\documentclass[11pt]{article}
\usepackage{amsmath, amssymb, graphics}

\usepackage[nosort]{cite}

\usepackage{amsbsy}
\usepackage{amsfonts}
\usepackage{amsmath}
\usepackage{graphicx}
\usepackage{array}
\usepackage{booktabs}
\usepackage{float}
\usepackage[usenames, dvipsnames]{color}
\usepackage{comment}
\usepackage{hyperref}
\numberwithin{equation}{section}

\usepackage{appendix}
\usepackage[nodayofweek]{date time}

\newcommand{\mathsym}[1]{{}}
\def\id{\protect{{1 \kern-.28em {\rm l}}}}

\def\be{\begin{eqnarray}}
\def\ee{\end{eqnarray}}

\makeatletter
\renewcommand\section{\@startsection {section}{1}{\z@}%
                                   {-3.5ex \@plus -1ex \@minus -.2ex}%
                                   {2.3ex \@plus.2ex}%
                                   {\normalfont\large\bfseries}}
\renewcommand\subsection{\@startsection{subsection}{2}{\z@}%
                                   {-3.25ex\@plus -1ex \@minus -.2ex}%
                                   {1.5ex \@plus .2ex}%
                                   {\normalfont\normalsize\bfseries}}

\makeatother

\def \foot {\footnote}

\def \ha {{1 \over 2}}

\def \ci{\cite}
\def \N {{\mathcal N}}

\def \del{\partial}

\def\s{\sigma}

\def\ov{\over}

\def\l{\lambda}

\def\foot{\footnote}

\def\det{\hbox{det}}
\def \ci {\cite}

\def \ha {{1 \over 2}}

\def \fo { {1\ov 4}}

\def \l  {\lambda}

\def \N {{\mathcal N}}

\def \O  { O}   

\def \N {{\mathcal N}}

\def \m {\mu}

\def \la {\label}

\def \l {\lambda}

\def \sql {{\sqrt \l}}

\newcommand{\rf}[1]{(\ref{#1})}
\def \ov {\over}

\def\N{{\cal N}}

\def \ha{{1\ov 2}}

\def \no {\nonumber}

\def \del {\partial}

\def \la {\label}

\def \l {\lambda}

\def \sql {{\sqrt \l}}

\def \ov {\over}

\def \varpi {{\rm w}}

\def \del {\partial} 

\def \s {\sigma}

 \def \n {\nu}

 \def \sql {\sqrt{\lambda}}

\newcommand{\preprint}[1]{\begin{table}[t]  
             \begin{flushright}               
             {#1}                             
             \end{flushright}                 
             \end{table}}                     
\renewcommand{\title}[1]{\vbox{\center\LARGE{#1}}\vspace{5mm}}
\renewcommand{\author}[1]{\vbox{\center#1}\vspace{5mm}}
\newcommand{\address}[1]{\vbox{\center\em#1}}

\begin{document}
\begin{titlepage}
\vskip -2cm
\date{\today}

\preprint{PUPT-2526\\ Imperial-TP-AT-2017-06}
\vskip -2.5cm

\begin{center}

\title{
Half-BPS Wilson loop and AdS$_2$/CFT$_1$}

\author{Simone Giombi$^{a}$, Radu Roiban$^{b}$, Arkady Tseytlin$^{c,}$\footnotemark}
\footnotetext{Also at Lebedev  Institute, Moscow.}

\address{${}^a$Department of Physics, Princeton University, Princeton, NJ 08544, USA}
\address{${}^b$Department of Physics, 
The Pennsylvania State University,\\ University Park, PA 16802, USA}
\address{${}^c$Blackett Laboratory, Imperial College, London SW7 2AZ, U.K.}


\end{center}

\vskip 1cm

\abstract{We study correlation functions of local operator insertions on the 1/2-BPS Wilson line in ${\cal N}=4$ super Yang-Mills theory. 
These correlation functions are constrained by the 1d superconformal symmetry 
preserved by the 1/2-BPS Wilson line
and define a defect CFT$_1$ living on the line. 
At strong coupling, a set of elementary operator insertions with protected scaling dimensions correspond to 
fluctuations of the dual fundamental string in AdS$_5 \times S^5$  ending on the line at the boundary 
and can be thought of as light fields propagating on the AdS$_2$ worldsheet. 
We use AdS/CFT techniques to compute the tree-level AdS$_2$ Witten diagrams describing the strong coupling limit of the four-point 
functions of the dual operator insertions. Using the OPE, we also extract the leading strong coupling corrections 
to the anomalous dimensions of the ``two-particle" operators built out of elementary excitations. 
In the case of the circular Wilson loop, we match our results for the 4-point functions of a special type of scalar 
insertions to the prediction of localization to 2d Yang-Mills theory.}

\vfill

\end{titlepage}

\eject \tableofcontents

\setcounter{footnote}{0}

\bigskip
\bigskip

\def \la {\label} \def \fo {{1\ov 4}}
\def \ed   {\end{document}}
\def \m  {\mu}  \def \n {\nu} \def \N {{\cal N}} \def \ov {\over} 
\def \ads {\ AdS$_2$\ } \def \del  {\partial} 

\def \beg  {\begin{equation}}
\def \eeg {\end{equation}}
 \def \tet {\textstyle}

\section{Introduction}


In the ${\cal N}=4$ supersymmetric Yang-Mills theory, it is natural to consider Wilson loop operators 
that include couplings to the six scalars $\Phi^I$ 
in the theory \cite{Maldacena:1998im, Rey:1998ik}
\begin{equation}\la{1}
W = {\rm tr}P e^{\oint dt\left(i\dot x^{\mu}A_{\mu}+|\dot x|\theta^I \Phi^I\right)}\,,
\end{equation}
where $x^{\mu}(t)$ is a closed loop, and $\theta^I(t)$ is a unit 6-vector. For generic contour and scalar couplings, 
these operators are only locally supersymmetric, but special choices of $x^{\mu}$ and $\theta^I$ lead to 
Wilson loops preserving various fractions of the superconformal symmetry of the ${\cal N}=4$ SYM theory \cite{Zarembo:2002an,Drukker:2007qr}. 
The most supersymmetric operator is obtained by taking the contour to be an infinite straight line (or circle), and $\theta^I$ a constant 6-vector, 
corresponding to a fixed direction in the scalar space: in this case the Wilson loop is 1/2-BPS, i.e. it preserves 16 of the 32 supercharges of the 
superconformal group $PSU(2,2|4)$. Making the choice  $\theta^I\Phi^I=\Phi^6$, this 1/2-BPS straight line operator is given by
\begin{equation}\la{2}
W = {\rm tr}P e^{\int dt \left(i A_t+\Phi^6\right)}
\end{equation}
where we have identified  the Euclidean time  $x^0=t\in (-\infty,\infty)$ to be the line that defines the operator.  

In this paper we will be interested in the computation of correlation functions of local operators inserted along the straight Wilson line, defined 
as follows. Given some local operators $O_i(t_i)$ transforming in the adjoint representation of the gauge group, 
one can define the gauge invariant correlator \cite{Drukker:2006xg}
\begin{eqnarray}
\langle \langle O_1(t_1) O_2(t_2)\cdots O_n(t_n)  \rangle \rangle &\equiv&  
\langle {\rm tr}P\big [O_1(t_1)\ e^{\int dt(iA_t +\Phi^6)} \ O_2(t_2)\  e^{\int dt(iA_t +\Phi^6)} \cdots \ O_n(t_n)\ e^{\int dt(iA_t +\Phi^6)}  
\big] \rangle  \cr
&\equiv& \langle {\rm tr}P\big [O_1(t_1)O_2(t_2)\cdots O_n(t_n) e^{\int dt(iA_t +\Phi^6)}\big]\rangle \ .
\label{1d-corr} 
\end{eqnarray}
The $SU(N)$ indices are contracted with the Wilson lines joining the various points, making this a gauge invariant observable. 
Since the expectation value of the straight Wilson line is trivial, this definition satisfies $\langle \langle 1\rangle \rangle =\langle W \rangle =1$.  
More generally, one should normalize the correlator on the right-hand side by the expectation value of the Wilson loop without insertions (this is relevant 
in the case of the 1/2-BPS circular loop, which has a non-trivial expectation value \cite{Erickson:2000af, Drukker:2000rr, Pestun:2007rz}). 
Note that, since operator insertions are equivalent to deformations of the Wilson line \cite{Drukker:2006xg, Cooke:2017qgm}, the complete 
knowledge of the correlators (\ref{1d-corr}) would, at least in principle, allow to compute the expectation 
value of general Wilson loops which are deformations of the line or circle. 

To understand the structure of the correlators (\ref{1d-corr}), it is useful to recall the symmetries preserved by the 1/2-BPS Wilson line. 
First, it is clear that it preserves an $SO(5)$ subgroup of the $SO(6)_R$ $R$-symmetry that rotates the 5 scalars $\Phi^a$, $a=1,\ldots, 5$ 
that do not couple to the Wilson loop. In addition, it preserves an  $SO(2,1)\times SO(3)$ subgroup of the 4d conformal group $SO(2,4)$, 
where the $SO(3)$ corresponds to rotations around
the line, and the generators of $SO(2,1)$ correspond to dilatations, translation and special conformal transformation along the line. This $SO(2,1)$ 
is the $d=1$ conformal group. Together with the 16 supercharges preserved by the loop, the 
symmetries of 1/2-BPS Wilson lines form the $d=1$, $\N=8$  superconformal group $OSp(4^*|4)$.

It follows that operator insertions along the Wilson line are classified by their representations under the $OSp(4^*|4)$ symmetry. In particular, 
they are labelled by their scaling dimension $\Delta$, corresponding to a representation of $SO(2,1)$, and by a representation of 
the ``internal" (from the point of view of the line) symmetry group $SO(3)\times SO(5)$. The set of correlators 
(\ref{1d-corr}) are then constrained by the $d=1$ conformal 
symmetry in a way analogous to  
higher dimensional CFTs.  
They can be interpreted as  characterizing
a defect CFT$_1$ living on the Wilson line 
\cite{Drukker:2006xg,Sakaguchi:2007ba, Cooke:2017qgm}. 
This CFT$_1$ should then be fully determined by its spectrum of scaling dimensions and OPE 
coefficients.
Because the ``double-bracket" correlators (\ref{1d-corr}) satisfy all the usual properties of CFT correlation functions, we may often talk about 
the $O_i(t_i)$ as operators in a 1d CFT, without referring to their (non-local) origin in SYM theory.  

Among the possible operator insertions, a special role is played by a set of ``elementary excitations" that fall into a short representation of 
the $OSp(4^*|4)$ symmetry with 8 bosonic plus 8 fermionic operators, and have protected scaling dimensions. 
The bosonic operators are the  5 scalars $\Phi^a$  (with dimension $\Delta=1$)  that do not couple to the Wilson line, which have $\Delta=1$, and the components 
of the field strength ${\mathbb F}_{ti}\equiv iF_{ti}+D_i \Phi^6$ (with dimension $\Delta=2$)
 along   the directions $i=1,2,3$  transverse to the  line. 
${\mathbb F}_{ti}$      is also known as the displacement operator, which measures the change of the Wilson loop under deformations orthogonal to the contour 
(this can be defined for any defect in a 
CFT).\footnote{The fact that the displacement operator has protected dimension $\Delta=2$ for a line 
defect in a 4d CFT is a general result, and follows from a Ward identity for the breaking of translations in the directions orthogonal to the defect, 
see e.g. \cite{Billo:2013jda, Gaiotto:2013nva, Billo:2016cpy}.} 

The fact that these operators have protected scaling dimensions implies that 
their 2-point functions (in the sense of (\ref{1d-corr}))  computed in planar SYM theory 
take the exact form 
\begin{equation}
\begin{aligned}\la{4}
\langle\langle \Phi^a(t_1)\Phi^b(t_2) \rangle \rangle =\delta^{ab} \frac{C_{\Phi}(\lambda)}{t_{12}^2}\,,\qquad \qquad 
\langle\langle {\mathbb F}_{ti}(t_1){\mathbb F}_{tj}(t_2) \rangle \rangle = \delta_{ij}\frac{C_{{\mathbb F}}(\lambda)}{t_{12}^4}\,,
\end{aligned}
\end{equation}
where the 't Hooft coupling $\l$ dependence appears only  in the normalization factors. These are proportional to the so-called Bremsstrahlung function $B(\l)$
defined in \cite{Correa:2012at} 
\begin{equation} \la{5}
 C_{\Phi}(\lambda)=2B(\lambda)\ , \qquad \qquad C_{{\mathbb F}}(\lambda)=12B(\lambda)\ , \qquad \qquad 
B(\lambda)=\frac{\sqrt{\lambda } \, I_2(\sqrt{\lambda })}{4\pi ^2\,  I_1(\sqrt{\lambda })}\ , \end{equation}
with  the leading terms  at weak and strong coupling expansions being  explicitly 
$B(\l) = {\l\ov 16 \pi^2} -  {\l^2\ov 384 \pi^2}  + O(\l^3)$ and  $B(\l) = {\sql\ov 4 \pi^2} -  {3\ov 8\pi^2}  + O({1\ov \sql})$
\ci{Drukker:2011za,Correa:2012at}.

The three-point functions of these elementary bosonic excitations 
vanish by the $SO(3)\times SO(5)$ symmetry.
The  four-point functions are expected to be non-trivial
functions of the positions  $t_r$  (constrained by the 1d conformal symmetry as reviewed in Section 3 below)
and of the 
  coupling constant $\l$.  Little is  known  about their structure  apart   from   the leading 
perturbative term in  the four-point of ${\mathbb F}_{ti}$  computed in \cite{Cooke:2017qgm}.

 In this paper, we will compute these four-point functions 
at strong coupling using the string theory in AdS$_5 \times S^5$ dual to planar $\N=4$ SYM.  
At strong coupling, Wilson loops are related by duality to  open string minimal surfaces in AdS$_5$ ending
on the contour defining the loop operator at the boundary. In the 
case of the 1/2-BPS Wilson line (or circle), the relevant minimal surface is an 
AdS$_2$ embedded in AdS$_5$ (and 
sitting at a point on the $S^5$). 
The  fundamental  open string  stretched in AdS 
preserves the same $OSp(4^*|4)$ as the 1/2-BPS Wilson line (see e.g. \cite{Gomis:2006sb}). In particular, 
the 1d conformal group $SO(2,1)$ is realized as the isometry of AdS$_2$. 

As 
we will review in Section 2, expanding the string action in static gauge 
around the minimal surface  solution, one finds \cite{Drukker:2000ep} that the  AdS$_2$ multiplet 
 of fluctuations transverse to the string
includes  5 massless 
scalars $y^a$ corresponding to the $S^5$ directions, three massive scalars $x^i$ with $m^2=2$ corresponding to AdS$_5$  fluctuations, and 8 fermionic modes with $m^2=1$. 
It is then natural to identify these 8+8 excitations, which may be thought as fields living in AdS$_2$, with the elementary CFT$_1$ insertions 
described above
 \cite{Sakaguchi:2007ba,Faraggi:2011bb,Fiol:2013iaa}. 
Indeed, the standard relation $m^2=\Delta(\Delta -d)$  between AdS$_{d+1}$ scalar masses and the corresponding  
CFT$_d$  operator dimensions in the present case implies that the massless $y^a$ fields 
should  be dual to $\Delta=1$ operators in CFT$_1$, namely the scalars $\Phi^a$, while the three AdS$_5$ fluctuations $x^i$ with $m^2=2$ should be dual to
the field strength operators ${\mathbb F}_{ti}$ with $\Delta=2$.\foot{
The   spectrum of quadratic superstring  fluctuations 
is the same    as in the case of  ``non-relativistic limit"  of AdS$_5\times S^5$ superstring 
\ci{Gomis:2005pg}   and was  suggested \ci{Sakaguchi:2007ba} 
to be related via AdS$_2$/CFT$_1$  to the   $ OSp(4^*|4)$  invariant 
    $\N=8$   
superconformal quantum mechanics of \ci{Bellucci:2003hn}.}

In general, in AdS/CFT the closed superstring  vertex operators  are mapped to single-trace gauge 
invariant local operators in the SYM theory. Including the open-string sector (with  open strings ending at the boundary) 
one should be able to describe the gauge-invariant operators (\ref{1d-corr})  that correspond to insertions of general local operators
along the Wilson loop. In this paper we will  limit our considerations only to insertions corresponding to the operators  with 
protected  scaling dimensions,  that   should be dual to  ``light"  fields  on the AdS$_2$ string world-sheet as described above. 
It would be of course interesting to work out the strong coupling description of more general operator insertions 
such as, for instance, the insertion of $\Phi^6$ (the scalar field that couples to the 1/2-BPS Wilson line). It was 
suggested in \cite{Alday:2007he} that this type of insertion may develop a large dimension $\Delta \sim \lambda^{1/4}$ at strong coupling, corresponding 
to massive states of the open string with $m^2 \sim 1/\alpha' \sim \sqrt \lambda $.
Another possibility is that the insertion of $\Phi^6$ corresponds at large $\lambda$ to a ``two-particle" worldsheet bound state $\sim y^a y^a$ made of $S^5$ fluctuations, which is the lowest dimension singlet at strong coupling.\footnote{We thank Juan Maldacena for suggesting this possibility.} We will extract the scaling dimension of such two-particle state from the 4-point function of $y$-fluctuations, see eq.~(\ref{del-yy}) below. 

\begin{figure}
\begin{center}
\includegraphics[width=0.7\textwidth]{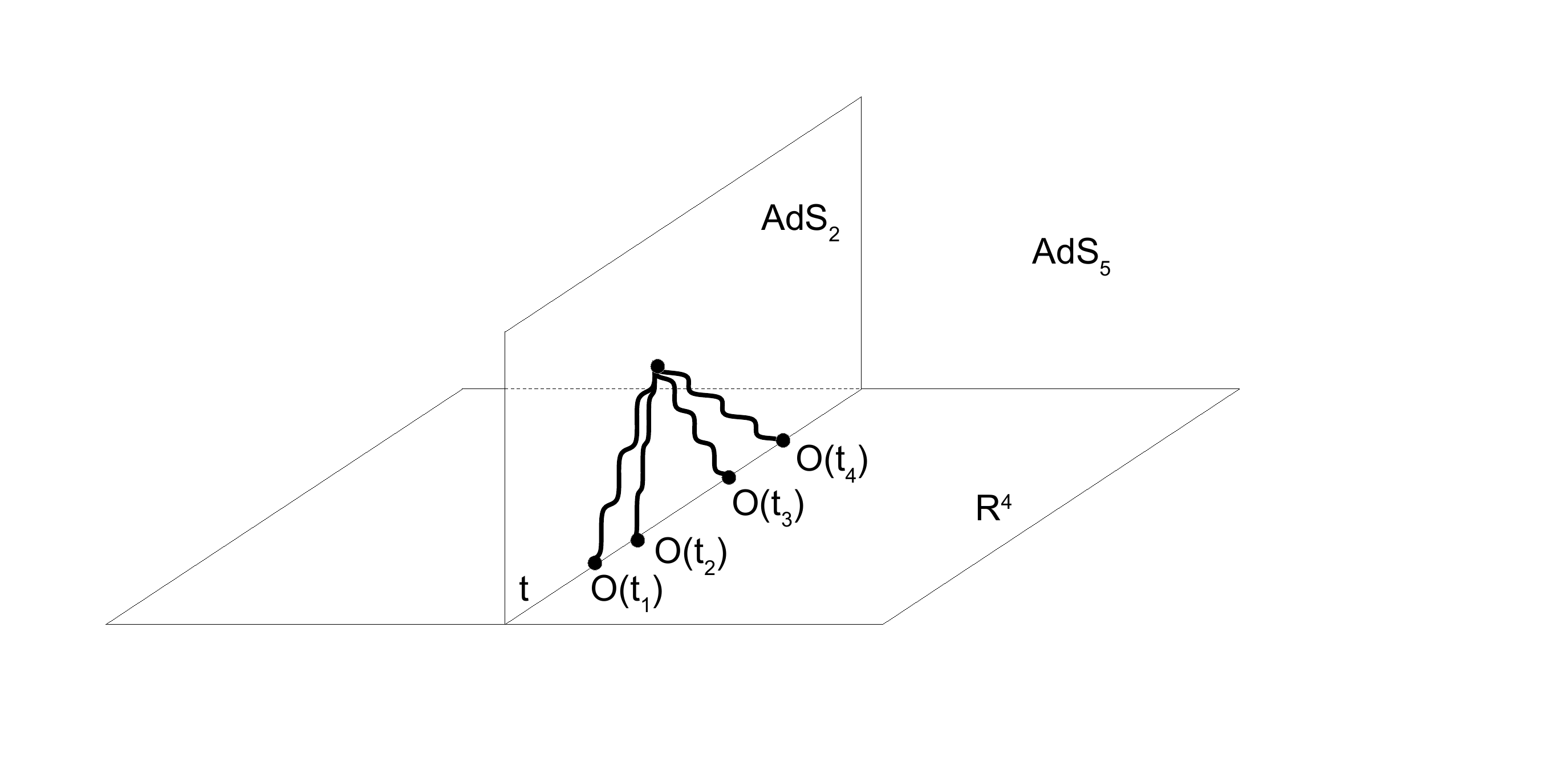}
\end{center}
\vskip -1cm
\caption{Four-point function of local operators inserted on the Wilson line from a Witten diagram on the AdS$_2$ worldsheet.}
\label{4ptAdS2}
\end{figure}

The expansion of the Nambu action around the classical solution yields the interaction vertices between the light fields. 
We will use these vertices to compute the corresponding tree-level Witten diagrams in AdS$_2$ and extract the strong coupling prediction for 
the four-point functions of the protected insertions on the Wilson line. This is depicted schematically in Figure \ref{4ptAdS2}. 
The calculation is similar to those in \ci{Liu:1998ty,Freedman:1998bj}, 
however,  we emphasize that the interpretation is different. In the supergravity calculations of \ci{Liu:1998ty,Freedman:1998bj}, one computes 
correlation functions of single-trace local operators, dual to closed string states, and the expansion parameter is $1/N^2$. In our case, 
we compute the correlators (\ref{1d-corr}) of insertions on the Wilson line, and the expansion parameter for the AdS$_2$ 
Witten diagrams is the inverse string tension, or $\frac{1}{\sqrt{\lambda}}$. Note that the 2d theory defined by the fundamental string 
action is expected to be UV finite, and thus the duality with the 1d CFT at the boundary should hold for any value of the 
coupling. In particular, the calculation of AdS$_2$ Witten diagrams involving loops should be well 
defined here.\footnote{For instance, by computing loop corrections 
to the boundary-to-boundary propagator one can verify that the elementary excitations are protected, as well 
as check the strong coupling expansion of the function $B(\lambda)$.}
%

Note also
that the AdS$_2$ worldsheet is not decoupled from the rest of the AdS$_5\times S^5$ bulk. For instance, one can consider processes where the 
worldsheet interacts with closed string modes propagating from the worldsheet to a point on the boundary away from the line 
(note, however, that these processes are suppressed in the large $N$ limit). This corresponds 
to correlation functions such as $\langle W {\rm tr}Z^J\rangle$ 
\cite{Berenstein:1998ij, Semenoff:2001xp,Pestun:2002mr} (and more generally one may consider mixed correlators 
of defect operators and operators inserted away from the defect). The picture is similar to the 
one discussed in \cite{Karch:2000ct, DeWolfe:2001pq, Aharony:2003qf} 
(see also, e.g., \cite{Billo:2013jda,Gaiotto:2013nva,Billo:2016cpy, Liendo:2016ymz, Rastelli:2017ecj} for recent related work) where one considers 
an AdS$_d$ brane inside AdS$_{d+1}$, and there is a defect CFT  living at the boundary of AdS$_d$. In our case, we have an  AdS$_2$ worldsheet inside AdS$_5$, and a codimension 3 defect in CFT$_4$ 
  (the Wilson line) at the boundary of AdS$_2$.\footnote{One may also consider the D3 and D5 branes dual to 1/2-BPS Wilson loops 
in higher-rank symmetric and antisymmetric representations \cite{Drukker:2005kx, Yamaguchi:2006tq, Gomis:2006sb, Gomis:2006im}: 
these branes have AdS$_2$ $\times$ $S^2$ 
or AdS$_2$ $\times$ $S^4$ worldvolumes, and preserve the same $OSp(4^*|4)$ symmetry as the fundamental string. Computing AdS$_2$ Witten diagrams 
in this case (after KK reduction on the sphere factors) should yield the correlators (\ref{1d-corr}) where the trace is taken to be in the 
symmetric or antisymmetric representations of rank $k\sim N$.}

Using the OPE expansion, we can also extract from the tree-level four-point functions the leading strong coupling corrections to the scaling dimensions of the ``two-particle" operators built of products of two of the protected insertions (with an arbitrary number of $t$-derivatives in between). For 
instance, we find that the $SO(3)\times SO(5)$ singlet operator with no derivatives built of scalar insertions has the dimension
\begin{equation}
\Delta_{y^a y^a} = 2-\frac{5}{\sqrt{\lambda}}+\ldots \,.
\label{del-yy}
\end{equation}
This is the lowest dimension unprotected operator in the spectrum at strong coupling.  
Let us again emphasize that these are not scaling dimensions of gauge invariant local operators in the ${\cal N}=4$ SYM theory but  are scaling dimensions of operator insertions on the Wilson line, defined as
in \rf{1d-corr}, \rf{4}
\begin{equation}\la{7}
\langle \langle O(t_1) O(t_2)\rangle \rangle = \langle {\rm tr}P[O(t_1)O(t_2)\ e^{\int dt(iA+\Phi^6)}]\rangle = \frac{C_{OO}}{t_{12}^{2\Delta_O}}\ , 
\end{equation}
with $O(t)$ being in the present case the operator dual to $y^a y^a$. In principle, the spectrum of dimensions 
of operators inserted on the Wilson line should be accessible 
 from the TBA approach  of \cite{Drukker:2012de, Correa:2012hh}, and 
it would be interesting to  reproduce  our  results in  this integrability-based  framework.\footnote{It would also be interesting to use integrability to reproduce a weak coupling Feynman graph approach to dimensions of operator insertions on the Wilson line.}
 More broadly, it would be important  to see how integrability is reflected in the structure of 
the Witten diagrams one computes in the AdS$_2$ worldsheet theory, perhaps uncovering an analog of the factorization of the S-matrix in integrable theories in flat space.  

While we focus on the straight line for most of the paper, our results can be also mapped to the circle by a (large) conformal transformation, 
as explained in Section 6. For a particular class of $S^5$ insertions on the circular loop that are expected to be captured by localization 
\cite{Drukker:2007yx,Giombi:2009ds, Pestun:2009nn, Giombi:2012ep}, we show in Section 6 that the result of the Witten diagram calculation in AdS$_2$ 
precisely matches the exact prediction derived from localization  to   2d YM theory.


As was appreciated in recent  discussions  of AdS$_2$/CFT$_1$   in the context of 
  dilaton-gravity models \ci{Almheiri:2014cka,
Maldacena:2016hyu,Maldacena:2016upp,Engelsoy:2016xyb,Jensen:2016pah} 
one can think of a system in AdS$_2$ as  having  asymptotic
1d reparametrization symmetry that is spontaneously broken down to  $SO(2,1)$,
which is the  isometry of AdS$_2$  metric. 
In our present case  the original definition of the  Wilson loop \rf{1}  has a reparametrization 
invariance which is fixed by the identification $x^0=t$ in \rf{2}, 
and the remaining conformal symmetry is the $SO(2,1)$ subgroup  of the 4d  conformal group
  that preserves the line. It is important to stress that compared to  the gravitational AdS$_2$ models 
in \ci{Almheiri:2014cka,Maldacena:2016hyu,Maldacena:2016upp,Engelsoy:2016xyb,Jensen:2016pah}
our bulk action \rf{2.4} is defined in fixed AdS$_2$ background, i.e. 
does not contain  gravity: before fixing the static gauge\foot{Defining the Wilson loop expectation  value
  in string theory in conformal gauge where one has two more  (compared to physical 
  static gauge) 
   dynamical coordinates  and ghosts one  would  end  effectively   with an integral over boundary  reparametrizations 
   (see \ci{Alvarez:1982zi,Rychkov:2002ni,Kristjansen:2012nz}). In this  case the identification between 
   the  operators  on the Wilson line  on the  gauge theory  side and  the string excitations 
   appears to become more intricate. This question deserves further  study.} 
the string action \rf{2.1} is reparametrization invariant, but gravity never becomes dynamical 
in critical superstring theory.
 In line with this, the boundary theory has no analog 
of the pseudo-Goldstone mode \cite{Maldacena:2016hyu} related to the (spontaneously broken) reparametrizations. 


\section{ AdS$_5 \times S^5$  string action   in static gauge  as AdS$_2$ bulk theory  action }


The   bosonic part of the superstring  action in AdS$_5 \times S^5$   has   the standard form 
\beg\la{2.1} 
S_B = \ha T   \int d^2\s \sqrt{h}\,   h^{\m\n} \Big[  \frac{1}{z^2}  \left(\partial_\m x^r\partial_\n x^r+\partial_\m z\partial_\n z\right)
+   {\partial_\m y^a\partial_\n y^a \ov (1+\fo  y^2)^2} \Big]  \ , \qquad  \ \ \  T= {\sqrt \lambda \ov 2 \pi}\ , 
\eeg
where $\s^\m= (t,s) $ are  Euclidean world-sheet  coordinates, 
$r=(0,i)=(0,1,2,3)$  label  coordinates of the Euclidean 4-boundary  and $a=1,...,5$ are $S^5$  labels. 
The minimal   surface  corresponding to  the straight Wilson line at the boundary is described by 
\beg \la{2.2}   z= s \ , \ \ \ \ \ \ \  x^0 = t \ , \ \ \ \ \   \ \ \    \qquad x^i=0\   , \ \ \ \  y^a=0 \ . \eeg 
The corresponding induced metric  is that of AdS$_2$, i.e.    $g_{\m\n} d\s^\m d\s^\n=   {1\ov s^2} ( dt^2 + ds^2)$. 

We 
will study correlators of   small fluctuations  of  ``transverse" 
string coordinates $(x^i, y^a)$ near this minimal surface
that will  thus propagate  in the  induced AdS$_2$ metric. The resulting global symmetry of the bosonic action will thus be 
$SO(2,1) \times [SO(3) \times  SO(6)]$.   To make the $SO(2,1)$ symmetry (which   will be 
 the conformal   symmetry at the corresponding  1d boundary theory)  manifest it is   useful to choose 
 the AdS$_2$ adapted  coordinates  and fix   the static gauge in which $z$ and $x^0$  do not fluctuate. 
The relevant embedding of AdS$_2$ into AdS$_5$  is described by  ($x^2\equiv x^i x^i  , \ \ i=1,2,3$)
\beg  \la{2.3} 
ds^2_5 =\frac{ (1+\fo  x^2)^2}{(1-\fo x^2)^2} ds^2_{2}  + \frac{dx^i dx^i}{(1-\fo x^2)^2} \ , \ \ \ \ \ \  \ \ \ \ \ \
ds^2_2 =\frac{1}{z^2} (dx_0^2+dz^2)      \ . 
\eeg
Starting with the  Nambu action and  fixing the static gauge by  the conditions on $x_0$ and $z$  as in \rf{2.2}
we  get 
\beg \la{2.4} 
S_B =  T \int d^2\s   \sqrt{\det \Big[   \frac{ (1+\fo  x^2)^2}{(1-\fo x^2)^2} \,  g_{\m\n} (\s)  + \frac{\del_\m x^i \del_\n x^i  }{(1-\fo x^2)^2}
 +   {\partial_\m y^a\partial_\n y^a \ov (1+\fo  y^2)^2}  \Big] } \equiv T \int d^2\s   \sqrt{ g}\  L_B   \ , 
 \eeg
 where $ g_{\m\n}= {1\ov s^2} \delta_{\m\n}$ is the  background AdS$_2$ metric. 
 This  action   can be interpreted as that of a  straight fundamental   string in AdS$_5 \times S^5$ 
 stretched   along  $z$, i.e. from the boundary towards the  center of AdS$_5$. 
 It  may be also viewed  as a 2d  field   theory of   3+5   scalars in AdS$_2$ geometry 
 with manifest  symmetry $SO(2,1) \times [SO(3) \times  SO(6)]$.
 Interpreted as a 2d  bulk  AdS$_2$  theory, it should thus have a   CFT$_1$  dual 
 living at  the $z=s=0$ boundary. As explained in the Introduction, this CFT$_1$ can be viewed as the
 defect CFT defined by operator insertions on the straight Wilson line.

 Expanding this action in powers of  $x^i$ and $y^a$ we get  
 \begin{align}   L_B =&\  L_2  + L_{4x}   + L_{2x,2y} + L_{4y}   + ...   \ , \la{2.5}\\
L_2=&\tet  \frac{1}{2}  g^{\m\n}\del_\m x^i \del_\n x^i  +  x^i x^i + \frac{1}{2} g^{\m\n}\del_\m y^a \del_\n y^a\ ,  \la{2.6} 
\\[2pt]
L_{4x} =&\ \tet  \frac{1}{8} (g^{\m\n}\del_\m x^i \del_\n x^i )^2  
                 - \frac{1}{4} (g^{\m\n} \del_\m x^i \del_\n x^j) \; (g^{\rho\kappa} \del_\rho x^i \del_\kappa  x^j)
\nonumber     \\
             &\ \tet + \frac{1}{4}  x^i x^i  (g^{\m\n} \del_\m x^j \del_\n x^j) + \frac{1}{2} x^i x^i\, x^j x^j \ , 
\label{2.7}            
\\[2pt]
L_{2x,2y}=&\ \tet  \frac{1}{4} (g^{\m\n}\del_\m x^i \del_\n x^i )\,(g^{\rho\kappa} \del_\rho y^a \del_\kappa  y^a) 
          - \frac{1}{2}  (g^{\m\n} \del_\m x^i \del_\n y^a) \; (g^{\rho\kappa} \del_\rho x^i \del_\kappa  y^a)\ , \la{2.8}
\\[2pt]
L_{4y} =&\  \tet
-\frac{1}{4} (y^b y^b) (g^{\m\n} \del_\m y^a \del_\n y^a) 
+\frac{1}{8}  (g^{\m\n}\del_\m y^a \del_\n y^a)^2
-\frac{1}{4}  (g^{\m\n} \del_\m y^a \del_\n y^b) \; (g^{\rho\kappa} \del_\rho y^a \del_\kappa  y^b)\ . \la{2.9}
\end{align}
Thus $x^i$ are 3  massive  ($m^2=2$) and  $y^a$  are  5  massless  scalars propagating  in AdS$_2$.

One may  also  include the fermionic  terms  coming   from  the corresponding 
AdS$_5 \times S^5$   superstring action as  in   \ci{Drukker:2000ep} (there will also be eight  2d fermions with mass 1). 
The resulting  2d  theory 
should be UV   finite  and  thus  should   be dual to a
quantum 1d CFT at the boundary  for any value of   the coupling $T={\sql \ov 2\pi} $. 
The coefficients in  the   correlation functions  computed in perturbation theory  will be   given by power series in $1\ov \sql$. 

At strong coupling ($\l \gg 1$) the 
correlators (\ref{1d-corr}) are expected to be reproduced by the AdS$_2$ amplitudes in the (super) string sigma model theory \rf{2.4}, 
with  the operators $O$ corresponding to particular string coordinates $X$, i.e. 
\begin{equation} 
\langle \langle O(t_1) O(t_2) ...O(t_n)  \rangle \rangle  ={ \langle   X(t_1) X(t_2) .... X(t_n) \rangle}_{_{\rm AdS_2}}  \ , \la{2.11} \
 \end{equation}
 where $\langle  ... \rangle{_{\rm AdS_2}} $  is the  expectation value in the  2d theory \rf{2.4}  
  corresponding  to Witten diagrams    with bulk-to-boundary  propagators  attached to the points  $t_1, ..., t_n$ at the boundary. 
As discussed in the Introduction,  the $X \sim  y^a$ in \rf{2.4} will correspond to the scalar 
 operators   $O \sim  \Phi^a$   ($a=1,...,5$)   of dimension $\Delta=1$ 
 while 
 $X \sim x^i$  will correspond to the generalized 
  field  strength components  $O\sim   {\mathbb F}_{it}$  with $\Delta=2$.
     
  The relation \rf{2.11}  can be understood as follows.
    The correlators $\langle \langle O(t_1) O(t_2) ...O(t_n)  \rangle \rangle $ in 
  \rf{1d-corr}   can be found by  first computing  a wavy-line  Wilson loop expectation value 
  $\langle W(C)\rangle$,  taking  functional derivatives over the contour function $C(t)$
  and then setting it to be a straight line. At weak coupling this procedure was 
  followed in   \cite{Cooke:2017qgm}.
  At   strong coupling $\langle W(C)\rangle$  is  assumed to be given  by the AdS$_5 \times S^5$ 
  open string path integral  with Dirichlet boundary conditions (implying that  disc-like or half-plane like 
  world-surface ends on a contour at the boundary of AdS$_5 \times S^5$). 
  To leading  order  in large  $\sqrt \lambda$ expansion   that   means   computing  the minimal area 
  of the  corresponding surface, i.e. the value of the (Euclidean)   string action on the  classical solution of the Dirichlet problem. 
  In the present  case of the  string action in the static gauge \rf{2.4}  interpreted as a   2d field theory in AdS$_2$ 
  this is equivalent to the  standard AdS/CFT procedure of computing the generating functional for 
  the  corresponding  CFT$_1$ boundary  correlators or $ { \langle   X(t_1) X(t_2) .... X(t_n) \rangle}_{_{\rm AdS_2}} $. Expanding the resulting on-shell  value 
  of the string action in powers of small deviations  of the boundary curve  from  the straight line will  then 
  give the correlators that   can  be equivalently  found  by computing the 
bulk Green's functions connected   to the boundary points by the   bulk-to-boundary propagators.
One  can check the  relation  \rf{2.11} explicitly  at the 2-point level  using the  wavy-line solution of 
\ci{Mikhailov:2003er}  (see also \ci{Kruczenski:2012aw}),  
reproducing the string  tree-level \ci{Mikhailov:2003er,Chernicoff:2009xc} 
and the  1-loop \ci{Buchbinder:2013nta}  corrections in the  strong-coupling expansion of 
the $B(\l)$ function in \rf{4},\rf{5}.

The Lagrangian \rf{2.5}  has no cubic terms, so the contribution to the simplest 4-point  tree-level correlation functions
of $x^i$ and $y^a$   will be  given just by the contact 4-point vertices in \rf{2.7}--\rf{2.9}. 
Below   we will compute the corresponding Witten diagrams  in AdS$_2$ connecting 
the 4-vertices to the boundary points  by   bulk-to-boundary   propagators as in, e.g.,   \ci{Liu:1998ty,Freedman:1998bj}.  
As we will be interested only in leading  large $\l$   (tree-level)   bosonic field   correlators 
 we will ignore the fermions. 

Note that while   we have made  a particular choice of AdS$_5$   coordinates in \rf{2.3}  the result  for the 
on-shell  AdS$_2$   amplitudes  (i.e. boundary operator   correlation functions)
should   be invariant under local field redefinitions (at least  in the case of 
 separated boundary points controlled by conformal invariance).


After  including  fermions and fixing  kappa-symmetry gauge  the superstring action
(generalizing \rf{2.1},\rf{2.4})  expanded near the 
1/2 BPS   straight line minimal surface   should  be describing a globally 
supersymmetric  field theory  \ci{Drukker:2000ep}   for  the   $OSp(4^*|4)$   multiplet of  8+8 bosons and fermions in AdS$_2$. Same symmetry   appears   on the dual gauge 
theory side.
 While in this paper we will discuss  only   4-point correlators of bosonic  coordinates, this  supersymmetry should allow  
also to determine the correlators   involving fermionic excitations.

\section{Four-point functions and conformal blocks in CFT$_1$}
\label{CFT1}

Before proceeding to  computation of correlators  of 2d fields in the AdS$_2$ theory \rf{2.4},\rf{2.5} let us make some 
general remarks about the structure of four-point functions in CFT$_1$.   

Local operators in a $d=1$ CFT  defined on a line $ \mathbb{R}=\{t\}$  which are 
 covariant under the conformal group $SO(2,1)$ 
are  labelled  just by their scaling dimension $\Delta$   (and possibly by some representation 
of an internal symmetry group which we suppress in this section).   
Let us consider the 4-point function of an operator $\O_{\Delta}(t)$.
The $SO(2,1)$  symmetry restricts the 4-point function to take the form
\begin{equation}
\langle \O_{\Delta}(t_1)\O_{\Delta}(t_2)\O_{\Delta}(t_3)\O_{\Delta}(t_4)\rangle = \frac{1}{(t_{12} t_{34})^{2\Delta}} 
\, {\cal G}(\chi)\ , \la{3.1}
\end{equation}
where $\chi \in (-\infty, \infty) $ is a conformally invariant cross ratio 
\begin{equation}
\chi = \frac{t_{12} t_{34}}{t_{13} t_{24}}\,.  \la{3.2} 
\end{equation}
Note that the usual cross ratios $u$, $v$ are not independent in $d=1$, i.e.  
\begin{equation}
u\equiv\frac{t_{12}^2 t_{34}^2}{t_{13}^2 t_{24}^2}=\chi^2\,,
\qquad \qquad 
v\equiv \frac{t_{14}^2 t_{23}^2}{t_{13}^2 t_{24}^2}= (1-\chi)^2 \   \la{3.3} \ . 
\end{equation}
 This is because the $SO(2,1)$ symmetry allows one  to fix three points 
on the line, leaving a single  free real parameter  as  the position of the fourth point. 
For example, if  we set $t_1=0$, $t_3=1$, $t_4=\infty$, then $\chi$ 
corresponds to the position $t_2$ of the second operator.  

\subsection{OPE expansion}

As in the case of higher dimensional CFT, the function ${\cal G}(\chi)$ in \rf{3.1} has an OPE expansion
\begin{equation}
{\cal G}(\chi) = \sum_{h} c_{\Delta,\Delta;h}\, \chi^{h}\  {}_2 F_1(h,h,2h,\chi) \ , 
\label{OPE-gen}
\end{equation}
where $h$ is the scaling dimension of the exchanged operator, $c_{\Delta,\Delta;h}=C_{\O_{\Delta}\O_{\Delta}O_h}^2/(C_{\O_{\Delta}\O_{\Delta}}^2 C_{\O_h \O_h})$ are normalized OPE coefficients, 
and $\chi^h\  {}_2 F_1(h,h,2h,\chi)$ is the exact conformal block in $d=1$ \cite{Dolan:2011dv}. 

We will also need the case of correlator of  operators with pairwise equal dimensions
\begin{equation}
\langle \O_{\Delta_1}(t_1)\O_{\Delta_2}(t_2)\O_{\Delta_1}(t_3)\O_{\Delta_2}(t_4)\rangle 
= \frac{1}{(t_{12}t_{34})^{\Delta_{1}+\Delta_2}}\,  \Big|\frac{t_{24}}{t_{13}}\Big|^{{\Delta_{12}}}\ {\cal G}(\chi)\,,\qquad 
\Delta_{12}\equiv \Delta_1-\Delta_2\ .
\label{pairwise}
\end{equation}
Here  the conformal block expansion reads  \cite{Dolan:2011dv}
\begin{equation}
{\cal G}(\chi) = \sum_h c_{\Delta_1,\Delta_2;h}\, \chi^h \  {}_2 F_1(h+\Delta_{12},h-\Delta_{12},2h,\chi)\,.
\label{pairwise-blocks}
\end{equation}
Note that in (\ref{pairwise}) we have written the result by choosing the $12\rightarrow 34$ channel (corresponding to $\chi\rightarrow 0$), 
which will be more convenient below. Of course, 
one may also write the 4-point function in the form
\begin{equation}\la{3.7} 
\langle \O_{\Delta_1}(t_1)\O_{\Delta_1}(t_2)\O_{\Delta_2}(t_3)\O_{\Delta_2}(t_4)\rangle 
= \frac{1}{t_{12}^{2\Delta_1}t_{34}^{2\Delta_2}} \ \tilde {\cal G}(\chi)\ , 
\end{equation}
where $\tilde {\cal G}(\chi)$ is related to ${\cal G}(\chi)$ in (\ref{pairwise}) by \ \ 
$\tilde {\cal G}(\chi) = \chi^{\Delta_1+\Delta_2}\,   {\cal G}(\chi^{-1})$. 

\subsection{Generalized free field OPE coefficients}
\label{gen-free}

It will be useful for what follows to collect some results for the OPE coefficients of generalized free fields 
(see,  e.g.,  \cite{Heemskerk:2009pn, Fitzpatrick:2011dm, Fitzpatrick:2012yx, Komargodski:2012ek, Gaiotto:2013nva}). 
In the case of the 4-point function 
of identical operators of dimension $\Delta$, the generalized free
 field 4-point function  has $ {\cal G}(u,v) = 1+ u^\Delta  + (u/v)^\Delta$, i.e. in $d=1$ \rf{3.3} 
 is given by 
\begin{equation}
\langle \O_{\Delta}(t_1)\O_{\Delta}(t_2)\O_{\Delta}(t_3)\O_{\Delta}(t_4)\rangle  = 
\frac{1}{(t_{12} t_{34})^{2\Delta}} \Big[1+\chi^{2\Delta}+\frac{\chi^{2\Delta}}{(1-\chi)^{2\Delta}} \Big] \ , \la{3.8} 
\end{equation}  
where we assumed unit normalization of the 2-point function. The operators exchanged in the OPE are just the identity and the 
tower of ``two-particle" operators 
\beg \la{3.9}
\big[\O_{\Delta}\O_{\Delta}\big]_{2n}\sim \O_{\Delta} \partial_t^{2n} \O_{\Delta} \eeg
 of dimension $2\Delta+2n$, $n=0,1,\ldots$. 
The corresponding OPE coefficients are given explicitly by
\begin{equation}
c_{\Delta,\Delta;2\Delta+2n} =
\frac{2\big[ \Gamma (2 n+2 \Delta )\big]^2 \Gamma (2 n+4 \Delta -1)}{\big[\Gamma (2 \Delta )\big]^2 \Gamma (2 n+1) \Gamma (4 n+4 \Delta -1)}\ , 
\label{OPE-gen-free1}
\end{equation}
as one can verify from the identity
\begin{equation}\la{3.11} 
\sum_{n=0}^{\infty} c_{\Delta,\Delta;2\Delta+2n}\, \chi^{2\Delta+2n} \ {}_2 F_1(2\Delta+2n,2\Delta+2n,4\Delta+4n,\chi) = 
\chi^{2\Delta}+\frac{\chi^{2\Delta}}{(1-\chi)^{2\Delta}} \,.
\end{equation}
While operators with odd number of derivatives do not appear in the OPE of identical $O_{\Delta}$'s, it will be useful for the case 
of operators carrying a flavor index (where odd $n$ can appear in the antisymmetric channel) to note the following result for 
the sum over odd $n$:
\begin{equation}
\sum_{n=0}^{\infty} c_{\Delta,\Delta;2\Delta+2n+1}\, \chi^{2\Delta+2n+1}  {}_2 F_1(2\Delta+2n+1,2\Delta+2n+1,4\Delta+4n+2,\chi) = -\chi^{2\Delta} +
\frac{\chi^{2\Delta}}{(1-\chi)^{2\Delta}} \,.
\label{OPEcoeff-odd-n}
\end{equation}
In  the case of pairwise identical operators, we have (cf. \rf{pairwise}) 
\begin{equation}
\langle \O_{\Delta_1}(t_1)\O_{\Delta_2}(t_2)\O_{\Delta_1}(t_3)\O_{\Delta_2}(t_4)\rangle = \frac{1}{t_{13}^{2\Delta_1} t_{24}^{2\Delta_2}}
= \frac{1}{(t_{12}t_{34})^{\Delta_{1}+\Delta_2}} \left|\frac{t_{24}}{t_{13}}\right|^{\Delta_{12}} \chi^{\Delta_1+\Delta_2}\,.
\label{free-pairwise}
\end{equation}
Here  the operators exchanged in the small $\chi$ expansion are $[\O_{\Delta_1}\O_{\Delta_2}]_n \sim \O_{\Delta_1}\partial_t^n \O_{\Delta_2}$ 
for all integer $n$ (both even and odd). The corresponding OPE coefficients are found to be 
\begin{equation}
c_{\Delta_1,\Delta_2;\Delta_1+\Delta_2+n} = 
\frac{(-1)^n \Gamma \left(n+2 \Delta _1\right) \Gamma \left(n+2 \Delta _2\right)\,  \Gamma \left(n+2 \Delta _1+2 \Delta _2-1\right)}{\Gamma \left(2 \Delta _1\right) \Gamma \left(2 \Delta _2\right)\,  \Gamma (n+1)\,  \Gamma \left(2 n+2 \Delta _1+2 \Delta _2-1\right)}\,.
\label{OPE-pairs}
\end{equation}
Indeed, one may verify that this agrees with the OPE expansion in (\ref{pairwise-blocks}) by checking that
\begin{equation}
\sum_{n=0}^{\infty} c_{\Delta_1,\Delta_2;\Delta_1+\Delta_2+n} \, \chi^{\Delta_1+\Delta_2+n} \ 
{}_2 F_1(\Delta_1+n,\Delta_2+n,2\Delta_1+2\Delta_2+2n,\chi) = \chi^{\Delta_1+\Delta_2}\,. \la{3.15}
\end{equation}

\section{Four-point function of $S^5$ fluctuations}
\label{4pt-S5}

In this section we compute the tree-level 4-point Witten 
diagram of the $S^5$ fluctuations $y^a$ in  the AdS$_2$   action in \rf{2.5}. 
As reviewed above, these are dual to the 5  SYM scalars $\Phi^a$, 
$a=1,\ldots,5$ (that do not appear in the exponent of the half-BPS Wilson line operator)
inserted along the line. The strong-coupling limit of the SYM correlator \ref{1d-corr} 
should be given by the tree-level  string coordinate amplitude as in \rf{2.11}. 

 By conformal symmetry, the 4-point function  should take the form  \rf{3.1}, i.e. 
\begin{equation}\langle y^{a_1}(t_1)y^{a_2}(t_2)y^{a_3}(t_3)y^{a_4}(t_4)\rangle_{{\rm AdS}_2} 
=
\langle\langle \Phi^{a_1}(t_1)\Phi^{a_2}(t_2)\Phi^{a_3}(t_3)\Phi^{a_4}(t_4)\rangle\rangle = \frac{\big[C_{\Phi}(\lambda)\big]^2}{t_{12}^2 t_{34}^2} \ 
G^{a_1a_2a_3a_4}(\chi) \ , 
\label{4point-Phi}
\end{equation} 
where $\chi$ is the conformally invariant cross ratio \rf{3.2}.
  In writing (\ref{4point-Phi}), 
we used the fact that the operators
 $\Phi^a$ have protected dimension $\Delta=1$,  i.e.  that their exact two-point function is\foot{Once again, when  referring to operators of 1d CFT we   understand 
them as insertions on the Wilson line.} 
\begin{equation}\la{4.2}
\langle y^{a_1}(t_1)y^{a_2}(t_2)\rangle_{{\rm AdS}_2} =
\langle \langle \Phi^{a_1}(t_1)\Phi^{a_2}(t_2)\rangle \rangle = \delta^{a_1a_2}\frac{C_{\Phi}(\lambda)}{t_{12}^2}\,.
\end{equation}
In  (\ref{4point-Phi}) we factored out $\big[C_{\Phi}(\lambda)\big]^2$ so 
that in the OPE limit $\chi\rightarrow 0$ we have $G^{a_1a_2a_3a_4}(\chi) = \delta^{a_1a_2}\delta^{a_3a_4}+O(\chi)$. The two-point normalization  factor 
$C_{\Phi}(\lambda)$ is related to the Bremsstrahlung function defined in 
\cite{Correa:2012at, Correa:2012hh}.
 We can always absorb this 
factor in the normalization of the operators, and we will do so in the following by choosing a canonical form of the 
bulk-to-boundary  propagators. 

The function $G^{a_1a_2a_3a_4}(\chi)$  (which is also a non-trivial  function  of 
 the coupling $\l$) can be decomposed into its   $SO(5)$ singlet, symmetric traceless and antisymmetric parts,
\begin{equation}
\begin{aligned}
G^{a_1a_2a_3a_4}(\chi) &= \tet G_S(\chi) \delta^{a_1a_2}\delta^{a_3a_4}+G_T(\chi) \left(\delta^{a_1a_3}\delta^{a_2a_4}+\delta^{a_2a_3}\delta^{a_1a_4}
-\frac{2}{5}\delta^{a_1a_2}\delta^{a_3a_4}\right) \\
&~ \ \ \ \qquad\ \ \ \ \qquad \qquad +G_A(\chi) \left(\delta^{a_1a_3}\delta^{a_2a_4}-\delta^{a_2a_3}\delta^{a_1a_4}\right)\ . 
\label{channels1}
\end{aligned}
\end{equation}
At strong coupling, these functions are expected to have the expansion (working in perturbation theory)
\beg
\label{channels-expa}
G_{S,T,A} (\chi)=G_{S,T,A}^{(0)}(\chi)+\frac{1}{\sqrt{\lambda}}G_{S,T,A}^{(1)}(\chi)+\ldots\,.
\eeg
\begin{figure}
\begin{center}
\includegraphics[width=0.7\textwidth]{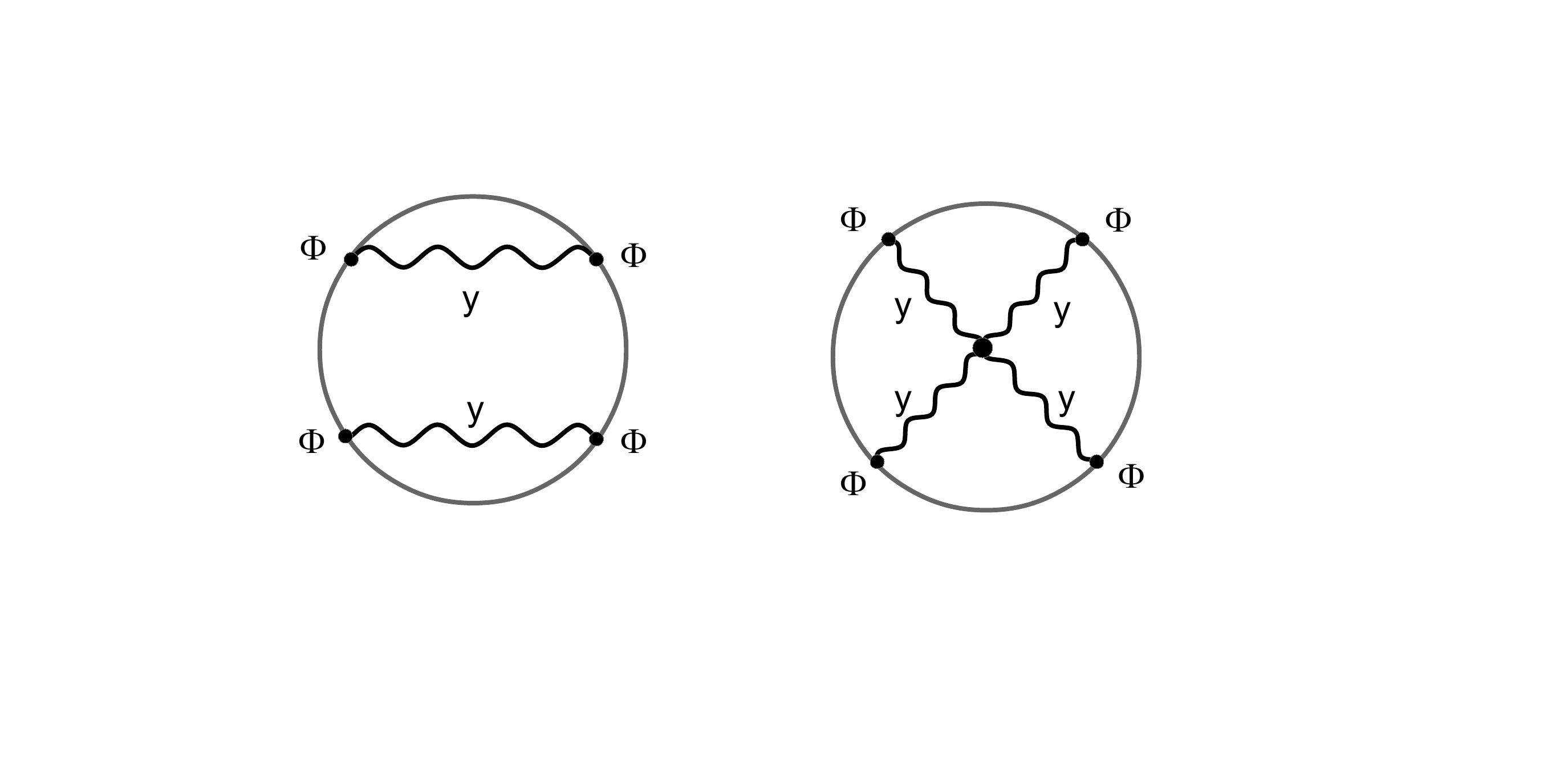}
\end{center}
\vskip -2cm
\caption{The disconnected and connected contributions to the 4-point function.}
\label{4pt-y}
\end{figure}
The leading terms here correspond to the disconnected contribution to the 4-point function, namely diagrams with two ``boundary-to-boundary" 
propagators, (see figure \ref{4pt-y}),
and are given by the generalized free field expression (cf. \rf{3.8})\footnote{Note that the separation between a connected and a disconnected 
contribution defined in (\ref{discon}) is natural from the point of view of the AdS$_2$ worldsheet perturbation theory, to all 
orders in $1\ov \sqrt{\lambda}$: in general, the disconnected contribution is given by a pair of loop-corrected boundary-to-boundary propagators. 
In the weak coupling limit, on the other hand, it is straightforward to see that 
the leading contribution in the planar limit is 
$\langle\langle \Phi^{a_1}(t_1)\Phi^{a_2}(t_2)\Phi^{a_3}(t_3)\Phi^{a_4}(t_4)\rangle\rangle 
=  \frac{\lambda^2}{64 \pi^2} \frac{1}{t_{12}^2t_{34}^2}
\left(\delta^{a_1a_2}\delta^{a_3a_4}+\frac{\chi^2}{(1-\chi)^2}\delta^{a_1a_4}\delta^{a_2a_3}\right) + O(\l^3) $, which is 
not exactly of the form (\ref{discon}), indicating that the connected contribution, defined from the point of 
view of the AdS$_2$ perturbation theory, should, in fact, 
contribute at leading order at small $\lambda$.
This 4-point function is the  leading free field term  in the large $N$ limit   with 
  the prefactor coming   from the   normalization of the 2-point function in \rf{4},\rf{5}   with 
$B(\l) = {\l \ov 16 \pi^2}  + O(\l) $. 
}
\begin{eqnarray}
\langle \langle \Phi^{a_1}(t_1)\Phi^{a_2}(t_2)\Phi^{a_3}(t_3)\Phi^{a_4}(t_4)\rangle \rangle_{\rm disconn.} =
\frac{\big[C_{\Phi}(\lambda)\big]^2}{t_{12}^2t_{34}^2}
\Big[\delta^{a_1a_2}\delta^{a_3a_4}+\chi^2 \delta^{a_1a_3}\delta^{a_2a_4}+\frac{\chi^2}{(1-\chi)^2}\delta^{a_1a_4}\delta^{a_2a_3}\Big]\,
\label{discon}
\end{eqnarray}  
which yields
\beg\la{4.6} 
G_S^{(0)}(\chi)=1+\tfrac{2}{5} G_T^{(0)}(\chi) \ ,  
\qquad G_T^{(0)}(\chi)=\tfrac{1}{2}\Big[\chi^2 +\frac{\chi^2}{(1-\chi)^2}\Big]\,,\qquad 
G_A^{(0)}(\chi)=\tfrac{1}{2}\Big[\chi^2 -\frac{\chi^2}{(1-\chi)^2}\Big]\,.
\eeg
The functions appearing at order $1\ov \sqrt{\lambda}$ in (\ref{channels-expa}) correspond to the leading contribution 
to the connected 4-point function at strong coupling, 
which comes from tree-level connected Witten diagrams.
These are  given by the  4-vertices in  \rf{2.9}  with four bulk-to-boundary propagators attached. 

We will adopt the following normalization of the   bulk-to-boundary  propagator (in  general dimension $d$) 
\begin{equation}
\begin{aligned}
K_{\Delta}(z,x;x') &= {\cal C}_{\Delta} \Big[\frac{z}{z^2+(x-x')^2}\Big]^\Delta \equiv  {\cal C}_{\Delta}\,  \tilde{K}_{\Delta}(z,x;x')\ , \\
{\cal C}_{\Delta} &=\frac{\Gamma\left(\Delta\right)}{2\pi^{d\ov 2}\Gamma\left(\Delta+1-{d\ov 2}\right)} \ . \la{4.7}
\end{aligned}
\end{equation}
In this normalization \cite{Penedones:2010ue, Fitzpatrick:2011ia}, the tree level 
two-point function of the dual boundary operator is $\langle \O_{\Delta}(x_1)\O_{\Delta}(x_2) \rangle ={ {\cal C}_{\Delta}\ov  x_{12}^{2\Delta}}$. \footnote{Note that this differs from the normalization adopted in \cite{Freedman:1998tz}, where 
${\cal C}_{\Delta}=\frac{\Gamma\left(\Delta\right)}{\pi^{d/2}\Gamma\left(\Delta-d/2\right)}$ was used. In that normalization, 
the two-point function of the dual operator is $\langle \O_{\Delta}(x_1) \O_{\Delta}(x_2) \rangle = {(2\Delta-d){\cal C}_{\Delta}\ov x_{12}^{2\Delta}}$.}
In the present case of $d=1$ and $\Delta=1$, we then have  ($t\equiv x^0$)
\begin{equation}
K_{\Delta=1}(z,t;t') = \frac{1}{\pi}\,  \frac{z}{z^2+(t-t')^2} \ , \qquad \qquad {\cal C}_{\Delta=1}={1\ov \pi} \ .   \la{4.8}
\end{equation}
When one has only  quartic contact diagrams (as in our present case), 
all tree-level 4-point functions  can be written in terms of the $D$-functions \cite{Liu:1998ty,DHoker:1999kzh, Dolan:2003hv} defined in the general  case of AdS$_{d+1}$ as 
\beg \la{4.9}
D_{\Delta_1\Delta_2\Delta_3\Delta_4}(x_1,x_2,x_3,x_4) = \int \frac{dz d^dx}{z^{d+1}} 
\tilde{K}_{\Delta_1}(z,x;x_1) \tilde{K}_{\Delta_2}(z,x;x_2)\ \tilde{K}_{\Delta_3}(z,x;x_3) \tilde{K}_{\Delta_4}(z,x;x_4)\,.
 \eeg
Note that derivatives in the vertices can be dealt with by using the identity (here $\del_\m=(\del_z,\del_r)$, \ $r=0,1, 2, ...,d-1$  and $g^{\m\n}= {z^2}\delta^{\m\n}$)
\begin{align}
&g^{\m\n}\partial_\m \tilde{K}_{\Delta_1}(z,x;x_1)\ \partial_\n\tilde{K}_{\Delta_2}(z,x;x_2)\no 
\\   & \qquad = 
\Delta_1\Delta_2
\left[\tilde{K}_{\Delta_1}(z,x;x_1)\tilde{K}_{\Delta_2}(z,x;x_2)-2x_{12}^2 \tilde{K}_{\Delta_1+1}(z,x;x_1)\tilde{K}_{\Delta_2+1}(z,x;x_2)\right]\ \la{4.10}\ .
\end{align}

\subsection{Connected part of the four-point function}

Returning  to our case of $d=1$,   let  us write the tree-level connected 4-point function \rf{4point-Phi} of $y^a$   coordinates
in \rf{2.5}  as
\begin{equation}
\langle\langle \Phi^{a_1}(t_1)\Phi^{a_2}(t_2)\Phi^{a_3}(t_3)\Phi^{a_4}(t_4)\rangle\rangle_{\rm conn} = 
\frac{2\pi}{\sqrt{\lambda}} \left({\cal C}_{\Delta=1}\right)^4\, Q^{a_1a_2a_3a_4}_{4y} \ , \la{4.11}
\end{equation}
where $Q_{4y}$ is obtained from the vertex $L_{4y}$ in \rf{2.9}. Explicitly, we find 
\begin{equation}
\begin{aligned}\la{4.12}
Q^{a_1a_2a_3a_4}_{4y}
=\ &  \Big[3 D_{1111}-2 t_{13}^2 D_{2121}-2 t_{14}^2 D_{2112}-2 t_{23}^2 D_{1221}
-2 t_{24}^2 D_{1212}\\ &\quad+4( t_{13}^2 t_{24}^2 +t_{14}^2 t_{23}^2-t_{12}^2 t_{34}^2)D_{2222}\Big]\delta^{a_1a_2}\delta^{a_3a_4}\\
+ &\Big[3 D_{1111}-2 t_{12}^2 D_{2211}-2 t_{14}^2 D_{2112}-2 t_{23}^2 D_{1221}-2 t_{34}^2 D_{1122}\\ &\quad 
+4( t_{12}^2 t_{34}^2+t_{14}^2 t_{23}^2-t_{13}^2 t_{24}^2)D_{2222}\Big]
\delta^{a_1a_3}\delta^{a_2a_4}\\
+ &\Big[3 D_{1111}-2 t_{12}^2 D_{2211}-2 t_{13}^2 D_{2121}-2 t_{24}^2 D_{1212}-2 t_{34}^2 D_{1122}\\  &\quad
+4( t_{12}^2 t_{34}^2+t_{13}^2 t_{24}^2-t_{14}^2 t_{23}^2)D_{2222}\Big]
\delta^{a_1a_4}\delta^{a_2a_3}\ .
\end{aligned}
\end{equation}
To write the result in a manifestly conformally invariant form, it is convenient to introduce the ``reduced" $\bar{D}$-functions that are functions of cross-ratios only. In general $d$, 
they are defined  in terms of \rf{4.9} as \cite{Dolan:2003hv}  ($\Sigma \equiv \frac{1}{2}\sum_i \Delta_i$)
\beg \la{4.15}
D_{\Delta_1\Delta_2\Delta_3\Delta_4}= 
\frac{\pi^{d\ov 2}\Gamma\left(\Sigma-{d\ov2}\right)}{2\, \Gamma\left(\Delta_1\right)\Gamma\left(\Delta_2\right)\Gamma\left(\Delta_3\right)\Gamma\left(\Delta_4\right)}
\frac{x_{14}^{2(\Sigma-\Delta_1-\Delta_4)} x_{34}^{2(\Sigma-\Delta_3-\Delta_4)}}
{x_{13}^{2(\Sigma-\Delta_4)} x_{24}^{2\Delta_2}}\bar{D}_{\Delta_1\Delta_2\Delta_3\Delta_4}(u,v)\ . 
\eeg
 $\bar{D}_{\Delta_1\Delta_2\Delta_3\Delta_4}$  
 can be written explicitly as the following Feynman 
parameter integral
\begin{equation}
\bar{D}_{\Delta_1\Delta_2\Delta_3\Delta_4}(u,v)=
\int d\alpha d\beta d\gamma\  \delta(\alpha+\beta+\gamma-1)\ 
\alpha ^{\Delta _1-1} \beta ^{\Delta _2-1} \gamma ^{\Delta _3-1} 
\frac{\Gamma \left(\Sigma-\Delta _4\right) \Gamma\left(\Delta_4\right)}
{\big(\alpha  \gamma + \alpha  \beta\, u  + \beta  \gamma\, v\big)^{\Sigma-\Delta _4}}\,.
\label{4.16}
\end{equation}
In $d=1$ where    $u=\chi^2$, $v=(1-\chi)^2$  we get 
$\bar{D}_{\Delta_1\Delta_2\Delta_3\Delta_4}$ as  a function of a single variable $\chi$. When 
the indices $\Delta_i$ are integers, the  integral \rf{4.16} 
can be evaluated explicitly. The basic example 
appearing in our calculations is 
\begin{equation}
\bar{D}_{1111}(\chi)=\frac{1}{\chi -1}\log \left(\chi^2\right)- \frac{1}{\chi}  \log\big[(1-\chi)^2\big]
\,.
\label{Db1-1d}
\end{equation}
One can check that this agrees with the  $d=1$  limit of the well-known result in general $d$
\begin{equation}\la{4.18}
\bar{D}_{1111}(u,v) = \frac{1}{z-\bar{z}}\Big[\log(z\bar{z})\log(\frac{1-z}{1-\bar{z}})+2{\rm Li}_2(z)-2{\rm Li}_2(\bar{z})\Big]\,,
\quad u = z\bar z\,,\quad v=(1-z)(1-\bar{z})
\end{equation}
after we set   $z=\bar z=\chi$  (cf. \rf{3.3}). The $\bar{D}$-functions with higher integer indices can be either evaluated directly using (\ref{4.16}), 
or expressed in terms of derivatives of $\bar{D}_{1111}$ using the identities listed in \cite{Dolan:2003hv}.

\def \logc {\log|\chi|}

Evaluating all the relevant integrals, the final result for \rf{4.11} takes the form 
\begin{equation}\la{4.19}
\langle\langle \Phi^{a_1}(t_1)\Phi^{a_2}(t_2)\Phi^{a_3}(t_3)\Phi^{a_4}(t_4)\rangle\rangle_{\rm conn} = 
\frac{({\cal C}_{\Delta=1})^2}{t_{12}^2 t_{34}^2} G^{a_1a_2a_3a_4}_{(1)}(\chi) \ , 
\end{equation}
where we factored out $({\cal C}_{\Delta=1})^2$ so that $ G^{a_1a_2a_3a_4}_{(1)}(\chi)$ corresponds to a canonical unit normalization. Separating out the singlet ($S$), symmetric 
traceless ($T$) and antisymmetric  ($A$) channels as in \rf{channels1},\rf{channels-expa}
\begin{equation}
\begin{aligned}
&G^{a_1a_2a_3a_4}_{(1)}(\chi) =\frac{1}{\sqrt{\lambda}}\Big{[}G_S^{(1)}(\chi)\,    \tet \delta^{a_1a_2}\delta^{a_3a_4}+G_T^{(1)}(\chi) \left(\delta^{a_1a_3}\delta^{a_2a_4}+\delta^{a_2a_3}\delta^{a_1a_4}
-\frac{2}{5}\delta^{a_1a_2}\delta^{a_3a_4}\right)\\
&~~~~~~~~~~~~~~~~~~~~~~~\qquad \qquad \qquad\qquad \quad 
+G_A^{(1)}(\chi) \left(\delta^{a_1a_3}\delta^{a_2a_4}-\delta^{a_2a_3}\delta^{a_1a_4}\right)\Big{]}
\label{channels}\ , 
\end{aligned}
\end{equation}
we find
\begin{align}
&G_S^{(1)}(\chi) = -\frac{2 \left(\chi ^4-4 \chi ^3+9 \chi ^2-10 \chi +5\right)}{5 (\chi -1)^2}
+\frac{\chi ^2 \left(2 \chi ^4-11 \chi ^3+21 \chi ^2-20 \chi +10\right)}{5 (\chi -1)^3}\log|\chi|\cr
&~~~~~~~~~\qquad \qquad \qquad \qquad ~~~~~~~~~~~~~~~~~~\ \ \ \ -\frac{2 \chi ^4-5 \chi ^3-5 \chi +10}{5\chi }\log|1-\chi|\ ,  \no \\
&G_T^{(1)}(\chi) = -\frac{\chi ^2 \left(2 \chi ^2-3 \chi +3\right)}{2(\chi -1)^2}
+\frac{\chi ^4 \left(\chi ^2-3 \chi +3\right)}{(\chi -1)^3}\log|\chi| -\chi ^3 \log|1-\chi| \ , \label{GSTA}   \\
&G_A^{(1)}(\chi) =\frac{\chi  \left(-2 \chi ^3+5 \chi ^2-3 \chi +2\right)}{2 (\chi -1)^2}
+\frac{\chi ^3 \left(\chi ^3-4 \chi ^2+6 \chi -4\right)}{(\chi -1)^3}\log|\chi|
-(\chi ^3-\chi ^2-1)\log|1-\chi|\nonumber 
\end{align}
Here and in what follows 
$\logc \equiv \ha \log (\chi^2) $  and $\log|1-\chi|\equiv  \ha \log \big[ (1-\chi)^2]$
where $\chi \in (-\infty, \infty)$.  Alternatively, we may  assume  that $\chi \in(0,1)$ (which, in particular, is 
sufficient for considerations of  the OPE below) and thus omit the absolute values, $\logc \to \log \chi $  and $\log|1-\chi|\to  
 \log (1-\chi)$. This is  sufficient  for  obtaining  the   expressions on the 
entire real 
 line using analytic continuation in $\chi$ (and crossing symmetry).

We can expand the  above  functions in the OPE limit $\chi \rightarrow 0$  as 
\begin{equation}
G_S^{(1)}(\chi) =\frac{1}{30} \chi ^2 \big(-60 \logc -43\big)+\frac{1}{30} \chi ^3 \big(-60 \logc-73\big)+ \frac{1}{60} \chi ^4 \big(-252 \logc -117\big)+\ldots \ , 
\label{Gsinglet}
\end{equation}
\begin{equation}\la{4.23}
G_T^{(1)}(\chi) = -\frac{3 }{2}\chi^2 -\frac{3 }{2}\chi ^3+\frac{1}{12} \chi ^4 \big(-36 \logc-18\big)+\ldots \ , 
\end{equation}
\begin{equation}\la{4.24}
G_A^{(1)}(\chi) =\frac{1}{6} \chi ^3 \big(24 \logc+7\big)+\frac{3}{4} \chi ^4 \big(8 \logc+5\big)+\ldots \ . 
\end{equation}
Since the term of order $\chi^2 \logc$ is 
absent from $G_T^{(1)}(\chi)$, this result implies that the symmetric 
traceless ``two-particle" operators $\Phi^{(a}\Phi^{b)}$ do not have an anomalous dimension. This is as expected since these operators, 
such as $Z^J$ with $Z=\Phi^1+i\Phi^2$ (inserted into the Wilson line)  
are BPS and hence protected \cite{Drukker:2006xg, Correa:2012hh}. 

On the other hand, the singlet $\Phi^a\Phi^a$ acquires an anomalous dimension due to the
presence of the  $\chi^2\logc$ term in (\ref{Gsinglet}). The same is true 
for other two-particle operators encoded in the higher powers $\chi^{2+n}\logc$. 
We will extract their scaling dimensions systematically in the next subsection. 

\subsection{Scaling dimensions of two-particle operators from OPE}
\label{OPE-sec}

Each of the functions $G_S(\chi)$, $G_T(\chi)$ and $G_A(\chi)$ in \rf{channels} 
is expected to have an OPE expansion of the form (\ref{OPE-gen}). To the leading order, where the 4-point function is 
given by the generalized free field expressions, the exchanged operators entering this expansion are the ``two-particle" 
operators of the form $\sim \Phi^a \partial_t^n \Phi^b$ 
(as  always   understood  as attached to the Wilson line), 
as reviewed in Section \ref{gen-free}. These can be decomposed in the irreducible representations of $SO(5)$
\begin{equation}
[\Phi \Phi]_{2n}^S\sim \Phi^a \partial_t^{2n}\Phi^a\,,\qquad 
[\Phi\Phi]_{2n}^{T}\sim \Phi^{(a}\partial_t^{2n}\Phi^{b)}\,, \qquad 
[\Phi\Phi]_{2n+1}^A \sim \Phi^{[a}\partial_t^{2n+1}\Phi^{b]}  \,. \la{4.39}
\end{equation}

The connected 4-point functions computed in the previous subsection encode the $1\ov \sqrt{\lambda}$ corrections 
to the scaling dimension of these operators, as well as the 
correction to the corresponding OPE coefficients. However, a difficulty arises in directly extracting this CFT data 
from the 4-point functions because of operator mixing.\footnote{We thank Marco Meineri and Carlo Meneghelli for useful discussions 
on these issues.} Due to degeneracies in the leading order two-particle spectrum, 
at the interacting level some of the operators in (\ref{4.39}) can mix with two-particle operators with the appropriate quantum 
numbers built out  of  generalized  gauge  field strength 
${\mathbb F}_{ti}$ or fermions (recall that the 8 fermions transform in the  $(2,4)$ representation 
of $SU(2)\times Sp(4)\simeq SO(3)\times SO(5)$). 
The singlet operators $[\Phi \Phi]_{2n}^S$ with $n>0$ can mix with ${\mathbb F}{\mathbb F}$ and two-fermion states, while the antisymmetric 
$[\Phi \Phi]_{2n+1}^A$ can mix with two-fermion states in the $(1,10)$ of $SU(2)\times Sp(4)$. 

Let us  start our analysis with the symmetric traceless channel. In this case, 
we expect that the corresponding operators $[\Phi\Phi]_{2n}^{T}$ should not be affected by mixing 
because there are no other two-particle operators with the same quantum numbers. We can write 
\begin{align}
&G_T(\chi) = \sum_h c_h \, \chi^h \,  F_h (\chi) 
= G_T^{(0)}(\chi)+\frac{1}{\sqrt{\lambda}}G_T^{(1)}(\chi)+\ldots \ , 
\label{OPE-GS}\\
& \ F_{h}(\chi)\equiv\ {}_2 F_1(h,h,2h,\chi)\,,  \la{4.26}
\end{align}
where the sum is over the primaries $[\Phi \Phi]_{2n}^T$ appearing in the OPE.  At large $\lambda$, we 
can write their dimension and  the OPE  coefficient as 
\begin{equation}
\begin{aligned}
h = 2+2n+\frac{1}{\sqrt{\lambda}} \gamma^{(1)}_{[\Phi\Phi]_{2n}^T}+\ldots \ , \ \ \qquad 
 \qquad c_h = c^{(0)}_{\Phi\Phi[\Phi\Phi]^T_{2n}}+\frac{1}{\sqrt{\lambda}}  c^{(1)}_{\Phi\Phi[\Phi\Phi]^T_{2n}}+\ldots \ . 
\label{exp-sql}
\end{aligned}
\end{equation}
Plugging the expansion (\ref{exp-sql}) into (\ref{OPE-GS}), we  get (see  \rf{4.6})
\begin{equation}
\begin{aligned}
\sum_{n=0}^{\infty} c^{(0)}_{\Phi\Phi[\Phi\Phi]^T_{2n}} \chi^{2+2n} F_{2+2n}(\chi) =  G_T^{(0)}(\chi) =\frac{1}{2}\Big[\chi^2 +\frac{\chi^2}{(1-\chi)^2}\Big] \ . \la{4.28}
\end{aligned}
\end{equation}
Comparing  this with the generalized free field result in (\ref{OPE-gen-free1}), the leading OPE coefficients are found to be 
\begin{equation}\la{4.29}
 c^{(0)}_{\Phi\Phi[\Phi\Phi]^T_{2n}} = \frac{\,  \big[\Gamma (2 n+2)\big]^2\ \Gamma (2 n+3)}{\,  \Gamma (2 n+1)\ \Gamma (4 n+3)}\,.
\end{equation}
From the terms of order $1\ov \sqrt{\lambda}$ in the expansions (\ref{exp-sql}), (\ref{OPE-GS}) we find the anomalous dimensions and corrections to the OPE coefficients. Expanding 
\beg\la{4.30}
\chi^h=\chi^{2+2n+{1\ov \sql}  \gamma^{(1)}+\ldots}=\tet  \chi^{2+2n}\big(1+\frac{1}{\sqrt{\lambda}}\gamma^{(1) }\logc+\ldots\big)\ ,  \eeg
 we see that the anomalous dimensions are determined by 
the $\logc$ terms in $G_T^{(1)}(\chi)$, via
\begin{equation}
\sum_{n=0}^{\infty} c^{(0)}_{\Phi\Phi[\Phi\Phi]^T_{2n}} \gamma^{(1)}_{[\Phi\Phi]^T_{2n}}  \chi^{2+2n} F_{2+2n}(\chi) = \big[G_T^{(1)}(\chi)\big]_{\logc} \ , 
\label{gam-rel}
\end{equation} 
where $ \big[G_T^{(1)}(\chi)\big]_{\logc} $ is the function multiplying $\logc$ in (\ref{GSTA}). This equation can be solved for any $n$ with the help 
of the orthogonality relation \cite{Heemskerk:2009pn}\footnote{SG is grateful to Vladimir Kirilin and Eric Perlmutter for many 
related discussions and collaboration on technically similar CFT calculations.}
\begin{equation}
\oint \frac{dz}{2\pi i} \  \frac{1}{z^2} \ 
z^{ \Delta+n} \ F_{\Delta+n}(z)\ z^{1-\Delta-n'} F_{1-\Delta-n'}(z)\  =\  \delta_{n,n'}\,, 
\label{ortho}
\end{equation} 
where the integral is over a contour around the origin in the complex plane. 
This result is valid for any $\Delta$, and can be verified,  for instance,  by using the 
series expansion of the hypergeometric function in \rf{4.26}. Using (\ref{ortho}), we get from (\ref{gam-rel})  
\begin{equation}
\gamma^{(1)}_{[\Phi\Phi]^T_{2n}} = \frac{1}{c^{(0)}_{\Phi\Phi[\Phi\Phi]^T_{2n}} }\oint  \frac{d\chi}{2\pi i}\  \chi^{-3-2n}F_{-1-2n}(\chi)\  \big[G_T^{(1)}(\chi)\big]_{\logc}  \,.\la{4.33}
\end{equation}
Evaluating the residue, we find that the result takes the remarkably simple form
\begin{equation}
\gamma^{(1)}_{[\Phi\Phi]^T_{2n}} = -2n^2-3n\,.\la{4.34}
\end{equation}
Thus  the strong-coupling expansion of  the scaling  dimension of the operator 
$O(t) = [\Phi \Phi]_{2n}^T\sim \Phi^{(a} \partial_t^{2n}\Phi^{b)}$ inserted in the Wilson line   as in (\ref{1d-corr})
is given by 
\begin{equation}
\Delta_{[\Phi\Phi]^T_{2n}} = 2+2n-\frac{2n^2+3n}{\sqrt{\lambda}}+O(\frac{1}{\lambda})\,.\la{4.36}
\end{equation} 
The vanishing of the anomalous dimension for $n=0$ reflects  the  fact that  the 
 operator $\Phi^{(a}\Phi^{b)}$ is protected. 
 The operators with $n>0$ are unprotected 
and belong to a long superconformal multiplet. 
Note that the anomalous dimension is negative for all $n>0$, indicating an effective attractive interaction between single-particle states.

Plugging the expansion (\ref{exp-sql}) into (\ref{OPE-GS}) results also in the following equation which determines the  leading  strong-coupling correction to the OPE coefficients 
\begin{equation}
\sum_{n=0}^{\infty} \chi^{2+2n}
\left[c^{(1)}_{\Phi\Phi[\Phi\Phi]^T_{2n}} F_{2+2n}(\chi)
+\frac{1}{2} c^{(0)}_{\Phi\Phi[\Phi\Phi]^T_{2n}}\gamma^{(1)}_{[\Phi\Phi]^T_{2n}}\partial_n F_{2+2n}(\chi)\right] = \big[G_T^{(1)}(\chi)\big]_{{\rm no}-\logc} \ , \la{4.37}
\end{equation}
where $\big[G_T^{(1)}(\chi)\big]_{{\rm no}-\logc}$  denotes the part of  (\ref{GSTA})
which does not  involve  the  $\logc$ term. 
It is straightforward to use this to extract $c^{(1)}_{\Phi\Phi[\Phi\Phi]^T_{2n}}$ for any given $n$. 
For example, the results for $n=0, 1,2$ including both 
order zero   \rf{4.29}  and $1\ov \sqrt{\lambda}$ correction are
\begin{equation}
\begin{aligned}\la{4.388}
c_{\Phi\Phi[\Phi\Phi]^T_{0}} =  1-\frac{3}{2 \sqrt{\lambda }}+\ldots \ , \quad 
c_{\Phi\Phi[\Phi\Phi]^T_{2}} = \frac{3}{5}-\frac{3}{20 \sqrt{\lambda }}+\ldots\,,\quad 
c_{\Phi\Phi[\Phi\Phi]^T_{4}} = \frac{5}{42}+\frac{335}{378 \sqrt{\lambda }}+\ldots
\end{aligned}
\end{equation}
For general $n$, we observe that the $O(\frac{1}{\sqrt{\lambda}})$ correction  to the OPE coefficients  in  \rf{exp-sql}   is given by the simple formula
\begin{equation}\la{4.377}
c^{(1)}_{\Phi\Phi[\Phi\Phi]^T_{2n}} = \frac{1}{2}\frac{\partial}{\partial n}
\left(c^{(0)}_{\Phi\Phi[\Phi\Phi]^T_{2n}}\gamma^{(1)}_{[\Phi\Phi]^T_{2n}}\right) \,.
\end{equation}
A relation of this type was found empirically in \cite{Heemskerk:2009pn} and proved in \cite{Fitzpatrick:2011dm} (see also \cite{Alday:2014tsa}). 
Explicitly, we get 
\begin{equation}\la{4.3777}
c^{(1)}_{\Phi\Phi[\Phi\Phi]^T_{2n}} =  {[ \Gamma(2n +2) ]^2 \ov \Gamma(4n +3)}   \big[ -( 3 + 34 n + 56 n^2 + 24 n^3)
+ 4n (n+1) ( 2n +1) ( 2n +3) ( H_{4n+3} - H_{2n} ) \big] 
  \,,
\end{equation}
where  $H_n=\sum^n_{k=1} {1 \ov k}$ is the harmonic number.

In the singlet and antisymmetric channels, the leading order OPE coefficients are determined by
\begin{equation}
\begin{aligned}\la{4.41} 
&1+\sum_{n=0}^{\infty} c^{(0)}_{\Phi\Phi[\Phi\Phi]^S_{2n}} \chi^{2+2n} F_{2+2n}(\chi) =  G_S^{(0)}(\chi) =1+\frac{1}{5}\Big[\chi^2 +\frac{\chi^2}{(1-\chi)^2}\Big]\ , \\
&\sum_{n=0}^{\infty} c^{(0)}_{\Phi\Phi[\Phi\Phi]^A_{2n+1}} \chi^{3+2n} F_{3+2n}(\chi) =  G_A^{(0)}(\chi) =\frac{1}{2}\Big[\chi^2 -\frac{\chi^2}{(1-\chi)^2}\Big]\ , 
\end{aligned}
\end{equation}
and  using   the results in Section \ref{gen-free}  are found to be
\begin{equation}
\begin{aligned}\la{4.42}
 c^{(0)}_{\Phi\Phi[\Phi\Phi]^S_{2n}} = \frac{2\big[\Gamma (2 n+2)\big]^2\  \Gamma (2 n+3)}{5\, \Gamma (2 n+1)\ \Gamma (4 n+3)}\,,\qquad \qquad 
 c^{(0)}_{\Phi\Phi[\Phi\Phi]^A_{2n+1}}=-\frac{\big[\Gamma (2 n+3)\big]^2\  \Gamma (2 n+4)}{\Gamma (2 n+2)\ \Gamma (4 n+2+3)}\,.
\end{aligned}
\end{equation}
In view of  the orthogonality relation (\ref{ortho}), we can extract the anomalous dimensions as 
\begin{equation}
\begin{aligned}\la{4.43}
&\gamma^{(1)}_{[\Phi\Phi]^S_{2n}} = \frac{1}{c^{(0)}_{\Phi\Phi[\Phi\Phi]^S_{2n}} }
\oint  \frac{d\chi}{2\pi i} \chi^{-3-2n}F_{-1-2n}(\chi) \ \big[G_S^{(1)}(\chi)\big]_{\log|\chi|}\ ,  \\
&\gamma^{(1)}_{[\Phi\Phi]^A_{2n+1}} = \frac{1}{c^{(0)}_{\Phi\Phi[\Phi\Phi]^A_{2n+1}} }
\oint  \frac{d\chi}{2\pi i} \chi^{-4-2n}F_{-2-2n}(\chi) \ \big[G_A^{(1)}(\chi)\big]_{\log|\chi|} \ . 
\end{aligned}
\end{equation}
Evaluating the residues we find, as in \rf{4.34}, 
simple quadratic polynomials in $n$ 
\begin{equation}
\gamma^{(1)}_{[\Phi\Phi]^S_{2n}} = -2n^2-3n-5\,,\qquad \qquad \gamma^{(1)}_{[\Phi\Phi]^A_{2n+1}} = -2n^2-5n-4\,.
\label{SA-gam}
\end{equation}
However, due to the mixing issues described above, these expressions  should be viewed as ``averages" of the anomalous dimensions over the 
operators appearing in the mixing (weighted by the corresponding OPE coefficients).\footnote{See \cite{Alday:2017xua} for a 
similar discussion in the context of  $1/N$  corrections to  
4-point functions of single trace operators in $\N=4$ SYM theory at strong coupling.} The exception 
is the singlet operator with $n=0$, which cannot mix with any other operator. In this case, 
(\ref{SA-gam}) yields
\begin{equation}
\Delta_{[\Phi\Phi]^S_{0}} = 2-{5\ov \sqrt{\lambda}}+O({1\ov \lambda})\,.
\end{equation}
Following similar  approach as  used   above in the  $[\Phi\Phi]^T_{2n}$ case, we can 
also extract the corresponding OPE coefficient
\begin{equation}
\la{4.38}
c_{\Phi\Phi[\Phi\Phi]^S_{0}} =  \frac{2}{5}-\frac{43}{30 \sqrt{\lambda }}+\ldots\,.
\end{equation}

It is interesting to notice  that in all of the  above expressions  \rf{4.36},\rf{SA-gam}  
 the large $n$ limit of the scaling dimensions has the same  asymptotic  form 
\beg \la{4.46} 
 \Delta_{n\gg 1} = 2n-{2n^2\ov \sqrt{\lambda}}+\ldots \ .
\eeg
We will find below that \rf{4.46}  is true  also  for the scaling dimensions extracted from 
the  mixed $x^2y^2$-correlators and $x^4$-correlators. Note that this implies that the perturbative result 
should not be trusted when $n$ becomes of order $\sqrt{\lambda}$, because then the leading term is comparable to 
the first perturbative correction (also, in such regime, 
contributions of massive string states should be already important, presumably corresponding to non-perturbative corrections 
to the 4-point function). Nevertheless, the form (\ref{4.46})  
is suggestive of a semiclassical limit with $n,\sqrt{\lambda}\gg 1$ and $\nu\equiv {n\ov \sqrt{\lambda}}$ fixed. 
Our results then suggest that in this limit  the dimensions  of such ``two-particle" operators have a universal  strong-coupling form 
\begin{equation}\la{4.47}
\Delta_n =\sqrt{\lambda}\,  f(\nu)\,,\qquad \qquad f(\nu) = 2\nu-2\nu^2+O(\nu^3)\,.
\end{equation}
This  behaviour may be captured by a semiclassical string calculation, analogous to the one in \cite{Drukker:2006xg, Miwa:2006vd, Sakaguchi:2007ea} 
where the insertions carried large $R$-charge, while here we just need large $SO(2,1)$ quantum number and no $R$-charge.

\section{Four-point functions  with  AdS$_5$  fluctuations}

Starting with the AdS$_2$ Lagrangian \rf{2.5}--\rf{2.9} we may also   compute   other four-point  correlators 
involving  AdS$_5$  coordinates $x^i$   which are dual to  the  dimension $\Delta =2$ operator 
${\mathbb F}_{it}$  inserted on the Wilson line.  Explicitly,  below we will compute  (cf. \rf{2.11})
\begin{align} 
&\langle x^{i_1}(t_1)x^{i_2}(t_2)y^{a_1}(t_3)y^{a_2}(t_4)\rangle_{{\rm AdS}_2} 
=\langle\langle\,   {\mathbb F}^{i_1}_{t}  (t_1)\,   {\mathbb F}^{i_2}_{t}  (t_2) \, \Phi ^{a_1}(t_3)\, \Phi^{a_2}(t_4)\, \rangle\rangle 
= \delta^{i_1 i_2}\delta^{a_1 a_2} \frac{G(\chi)}{t_{12}^4 t_{34}^2}  \ ,\la{5.1}\\
&\langle x^{i_1}(t_1)x^{i_2}(t_2)x^{i_3}(t_3)x^{i_4}(t_4)\rangle_{{\rm AdS}_2} 
=\langle\langle \,  {\mathbb F}^{i_1}_{t}  (t_1)\,   {\mathbb F}^{i_2}_{t}  (t_2) \,  {\mathbb F}^{i_3}_{t}  (t_3)\, 
{\mathbb F}^{i_4}_{t}  (t_4)\, \rangle\rangle 
=  \frac{G^{i_1i_2i_3i_4}(\chi)}{t_{12}^4 t_{34}^4}  \ .\la{5.2}
\end{align}
Since the Wilson line is 1/2-BPS these two correlators should be related to the correlation function of four $S^5$ fluctuations
by supersymmetry transformations.

\def \half {{\textstyle{1\ov 2}}}

\subsection{Two AdS$_5$  and two $S^5$ fluctuations}

The leading-order contribution to the connected part of  the  correlator  \rf{5.1}   may be written as (cf. \rf{4.7},\rf{4.8})
\beg
\frac{G_{\rm conn} (\chi)}{t_{12}^4 t_{34}^2} = \frac{2\pi}{\sql} 
({\cal C}_{\Delta=1}  {\cal C}_{\Delta=2})^2 Q_{xy} \ , \la{5.3}
\qquad \qquad   {G_{\rm conn} (\chi)}\equiv  {\cal C}_{\Delta=1}  {\cal C}_{\Delta=2}\,  G_{(1)} (\chi) \ ,
\eeg
where 
\begin{align}\la{5.4}
Q_{xy}= &  -\int \frac{dtds}{s^2}
\Big[
g^{\mu\nu}\partial_\mu{\tilde K}_2(t_1)\partial_\nu{\tilde K}_2(t_2)   g^{\rho\sigma}\partial_\rho {\tilde K}_1(t_3)\partial_\sigma {\tilde K}_1(t_4)\no
\\
&
-g^{\mu\nu}\partial_\mu{\tilde K}_2(t_1)\partial_\nu {\tilde K}_1(t_3) g^{\rho\sigma}\partial_\rho{\tilde K}_2(t_2)\partial_\sigma {\tilde K}_1(t_4)
-g^{\mu\nu}\partial_\mu{\tilde K}_2(t_1)\partial_\nu {\tilde K}_1(t_4) g^{\rho\sigma}\partial_\rho{\tilde K}_2(t_2)\partial_\sigma {\tilde K}_1(t_3)
\Big]
\nonumber
\\
=&\ 4  \Big(
      D_{2211} + 
   2t_{12}^2 D_{3311} - 
   2 t_{13}^2D_{3221}  - 
   2 t_{23}^2D_{2321}  - 
   2 t_{14}^2  D_{3212}  - 
   2  t_{24}^2   D_{2312}
   \cr
&\qquad\ \ \ \ \ \ 
   + 
   2 t_{34}^2 D_{2222}  + 
   4  t^2_{14}   t_{23}^2     D_{3322}  + 
   4 t_{13}^2 t_{24}^2  D_{3322} - 
   4 t_{12}^2 t_{34}^2 D_{3322} \Big)\ .
\end{align}
As a result, the  function $G_{(1)} (\chi) $ in \rf{5.3}
\beg\la{5.5}
G_{(1)}(\chi)=-\frac{4}{\sql}\,\Big[   1   -   \big(\half - \chi^{-1}\big)   \ln   |1 - \chi| \Big] \ . 
\eeg


Similarly to the discussion in Section \ref{OPE-sec}, we  may also  extract  the scaling dimensions of two-particle operators appearing in the OPE. 
In this case the  relevant operators  are 
\begin{equation}\la{5.6}
[\Phi^a {\mathbb F}_{it}]_n \sim \Phi^a \partial_t^n {\mathbb F}_{it}
\end{equation}
 that  have  dimension $3+n+O({1\ov \sqrt{\lambda}})$  and  correspond to mixed $xy$ two-particle states. Let us first rewrite the 4-point function  \rf{5.1},\rf{5.3}
by relabeling $t_2 \leftrightarrow t_3$ 
\begin{equation}
\langle \langle {\mathbb F}_{it}(t_1)\Phi^a(t_2) {\mathbb F}_{jt}(t_3)\Phi^b(t_4)\rangle \rangle_{\rm conn}
=\delta^{ab}\delta_{ij}\frac{{\cal C}_{\Delta=1} {\cal C}_{\Delta=2}}{(t_{12}^2t_{34}^2)^{3/2}}
\Big(\frac{t^2_{24}}{t^2_{13}}\Big)^{1/2}\, G^{(1)}_{xy}(\chi)\,,  \la{5.7}
\end{equation}
where from  \rf{5.5}   we get
\begin{equation}
G^{(1)}_{xy}(\chi) = \chi^3 G_{(1)}(\chi^{-1})=
-\frac{4\chi^3}{\sqrt{\lambda}} \Big[   1 + (\half - \chi)  \log \frac{|\chi|}{|1-\chi|}\ \Big]   \,.\la{5.8}
\end{equation}
The corresponding disconnected contribution appearing at leading order is (see (\ref{free-pairwise}))
\begin{equation}
G^{(0)}(\chi) 
= \frac{1}{(t_{12}^2t_{34}^2)^{3/2}} \Big(\frac{t_{24}^2}{t_{13}^2}\Big)^{{1}/{2}}\, \chi^{3}\,.
\end{equation}
Using (\ref{OPE-pairs}), this determines the leading order OPE coefficients appearing in the expansion (\ref{pairwise-blocks}) 
\begin{equation}\la{5.10}
c^{(0)}_{\Phi {\mathbb F}[\Phi {\mathbb F}]_n} = \frac{\Gamma (n+2) \, \Gamma (n+4)\,  \Gamma (n+5)}{6\,  \Gamma (n+1) \, \Gamma (2 n+5)}\,.
\end{equation}
To extract the anomalous dimensions, we can use the following generalization of the orthogonality relation (\ref{ortho})
\begin{align}\la{5.11}
&\oint \frac{dz}{2\pi i} \frac{1}{z^2} z^{\Delta+n} F_{\Delta+n,a}(z)z^{1-\Delta-n'} F_{1-\Delta-n',a}(z)= \delta_{n,n'}\,, \\
&F_{h,a}(z) \equiv {}_2 F_1(h+a,h-a,2h,  z)\ . \no
\end{align}
In our case, we need $\Delta=3$ and $a=1$, see (\ref{pairwise-blocks}). Then  the anomalous dimensions 
are given by
\begin{equation}
\gamma^{(1)}_{[\Phi {\mathbb F}]_n} = \frac{1}{c^{(0)}_{\Phi {\mathbb F}[\Phi {\mathbb F}]_n}}
\oint \frac{d\chi}{2\pi i} \  \chi^{-n-4} F_{-n-2,1}(\chi) \ 2 \chi ^3 (2 \chi -1)\ , \la{5.12}
\end{equation}
where we have used  that $\big[G^{(1)}_{xy}(\chi)\big]_{\log|\chi|}=2 \chi ^3 (2 \chi -1)$. Evaluating the residue, we find
\begin{equation}
\gamma^{(1)}_{[\Phi {\mathbb F}]_n} = -\frac{n^2}{2}-\frac{5 n}{2}-2\,.
\label{gam-PF}
\end{equation}
Let us separate the cases of even and odd $n$.  For even $n$  we expect that the operators $[\Phi {\mathbb F}]_{2n}$ can mix 
with two-fermion states in the same representation.\footnote{The product of two $(2,4)$ representations 
of $SU(2)\times Sp(4)$ contains the $(3,5)$ of $SO(3)\times SO(5)$. The vector of $SO(5)$ corresponds 
to the antisymmetric symplectic-traceless representation of $Sp(4)$.} For odd $n$, on the other hand, we do not expect mixing with two fermion states, and 
from (\ref{gam-PF}) we get 
\begin{equation}
\begin{aligned}\la{5.14}
&\Delta_{[\Phi {\mathbb F}]_{2n+1}}=4+2n-\frac{2n^2+7n+5}{\sqrt{\lambda}}+O(\frac{1}{\lambda})\,.
\end{aligned}
\end{equation}
For  large $n$, we recover the universal form \rf{4.46},\rf{4.47} found from the analysis of the $y$-correlators. 
Note that 
the dimension $\Delta_{[\Phi {\mathbb F}]_{2n+1}}$ in \rf{5.14} 
is  the same as $\Delta_{[\Phi\Phi]^T_{2n'}}$ in (\ref{4.36}) for $n'=n+1$. 
This is consistent with the fact that these operators should belong to the same long supermultiplet.

\subsection{Four  AdS$_5$  fluctuations}

Finally, let us compute the four-point function of the  three AdS fluctuations $x^i$ \rf{5.2} using  similar normalization 
for the connected part as in \rf{5.3}
\beg \la{5.15} 
  G^{i_1i_2i_3i_4}_{\rm conn}  (\chi) =({\cal C}_{\Delta=2})^2\, G_{(1)}^{i_1i_2i_3i_4}(\chi)\ , 
\eeg
with  (cf. \rf{channels})
\begin{align}
G_{(1)}^{i_1i_2i_3i_4}(\chi) = & 
\delta^{i_1i_2}\delta^{i_3i_4} G_S^{(1)}(\chi) + G^{(1)}_A(\delta^{i_1i_3}\delta^{i_2i_4}-\delta^{i_1i_4}\delta^{i_2i_3})
\cr
&\qquad \tet \qquad \qquad \ \ \ 
+ G_T^{(1)}(\delta^{i_1i_3}\delta^{i_2i_4}+\delta^{i_1i_4}\delta^{i_2i_3}-\frac{2}{3}\delta^{i_1i_2}\delta^{i_3i_4} )\ . \la{5.16}
\end{align}
The irreducible $SO(3)$  singlet, symmetric traceless and  antisymmetric parts   are found to be
\begin{align}
{G_S^{(1)}(\chi)}=&
-\frac{\left(24 \chi ^8-90 \chi ^7+125 \chi ^6-76 \chi ^5+125 \chi ^4-306 \chi ^3+438 \chi ^2-288 \chi +72\right)}{9
   (\chi -1)^4}
\cr
&
-\frac{2  \left(4 \chi ^6-\chi ^5-6 \chi +12\right)}{3 \chi }\log |1-\chi |
\nonumber\\[2pt]
&
+\frac{2  \chi ^4 \left(4 \chi ^6-21 \chi ^5+45 \chi ^4-50 \chi ^3+30 \chi ^2-6 \chi +2\right)}{3 (\chi -1)^5}\log | \chi |\ , 
\\
{G_T^{(1)}(\chi)}=&
-\frac{\left(48 \chi ^4-198 \chi ^3+313 \chi ^2-230 \chi+115\right) \chi ^4}{12 (\chi -1)^4}
-\frac{1}{2} (8 \chi -5) \chi ^4 \log | 1-\chi |
\nonumber\\[2pt]
&
+\frac{\left(8 \chi ^6-45 \chi ^5+105 \chi ^4-130 \chi ^3+90 \chi ^2-30 \chi+10\right) \chi ^4 }{2 (\chi -1)^5}\log |\chi |
\ , \\
{G_A^{(1)}(\chi)} =&
-\frac{(\chi -2) \left(48 \chi^6-90 \chi ^5+91 \chi ^4+4 \chi ^3-17 \chi ^2+18 \chi -6\right) \chi }{12 (\chi -1)^4}
\cr
&
-\frac{1}{2} \left(8 \chi ^5-3 \chi ^4+2\right) \log | 1-\chi |
\nonumber\\[2pt]
&
+\frac{(\chi -2) \left(8 \chi ^4-27 \chi^3+41 \chi ^2-28 \chi +14\right) \chi ^5 }{2 (\chi -1)^5}\log | \chi |\ . \la{5.19}
\end{align}
The two-particle states encoded in the OPE  of the 4-point function of $x$ fluctuations are 
\begin{equation}\la{5.20}
[{\mathbb F}{\mathbb F}]_{2n}^S \sim {\mathbb F}_{ti}\partial_t^{2n}{\mathbb F}_{it}\,, \qquad 
[{\mathbb F}{\mathbb F}]_{2n}^T \sim {\mathbb F}_{t(i}\partial_t^{2n}{\mathbb F}_{j)t}\,,\qquad
[{\mathbb F}{\mathbb F}]_{2n+1}^A \sim {\mathbb F}_{t[i}\partial_t^{2n+1}{\mathbb F}_{j]t}\,. 
\end{equation}
The calculation of their anomalous dimensions 
follows the same steps as outlined in the previous sections. The disconnected contributions to the 4-point function are 
(cf. \rf{4.6})
\begin{equation}
\la{5.21}
G_S^{(0)}(\chi)=1+\tfrac{2}{3}  G_T^{(0)}(\chi) \ , \qquad 
G_T^{(0)}(\chi)=\tfrac{1}{2}\Big[\chi^4 +\frac{\chi^4}{(1-\chi)^4}\Big]\,,\qquad 
G_A^{(0)}(\chi)=\tfrac{1}{2}\Big[\chi^4 -\frac{\chi^4}{(1-\chi)^4}\Big]\,,
\end{equation}
from which, using (\ref{OPE-gen-free1}), we find the leading OPE coefficients 
\begin{equation}
\begin{aligned}
&c^{(0)}_{{\mathbb F}{\mathbb F}[{\mathbb F}{\mathbb F}]_{2n}^S} = \frac{\big[\Gamma (2 n+4)\big]^2 \, \Gamma (2 n+7)}{54 \, \Gamma (2 n+1)\  \Gamma (4 n+7)}\ , \\
&c^{(0)}_{{\mathbb F}{\mathbb F}[{\mathbb F}{\mathbb F}]_{2n}^T} = \frac{\big[\Gamma (2 n+4)\big]^2\,  \Gamma (2 n+7)}{36 \, \Gamma (2 n+1)\  \Gamma (4 n+7)}\,,\qquad 
&c^{(0)}_{{\mathbb F}{\mathbb F}[{\mathbb F}{\mathbb F}]_{2n+1}^A} =-\frac{\big[\Gamma (2 n+5)\big]^2 \Gamma (2 n+8)}{36 \, \Gamma (2 n+2)\  \Gamma (4 n+9)}\ .\la{5.22}
\end{aligned}
\end{equation}
Then  starting with  the OPE    (\ref{OPE-gen}), expanding in powers of $1\ov \sqrt{\lambda}$ and using the orthogonality relation (\ref{ortho}), we find 
\begin{equation}
\begin{aligned}
&\gamma^{(1)}_{[{\mathbb F}{\mathbb F}]_{2n}^S} = \frac{1}{c^{(0)}_{{\mathbb F}{\mathbb F}[{\mathbb F}{\mathbb F}]_{2n}^S}}
\oint  \frac{d\chi}{2\pi i} \chi^{-5-2n}F_{-3-2n}(\chi)\ \big[ G_S^{(1)}(\chi)|_{\log|\chi|} = -2 n^2-7 n-2  \\
&\gamma^{(1)}_{[{\mathbb F}{\mathbb F}]_{2n}^T} =\frac{1}{c^{(0)}_{{\mathbb F}{\mathbb F}[{\mathbb F}{\mathbb F}]_{2n}^T}}
\oint  \frac{d\chi}{2\pi i} \chi^{-5-2n}F_{-3-2n}(\chi)  \ \big[G_T^{(1)}(\chi)\big]_{\log|\chi|}=-2n^2-7n-5 \\
&\gamma^{(1)}_{[{\mathbb F}{\mathbb F}]_{2n+1}^A} = \frac{1}{c^{(0)}_{{\mathbb F}{\mathbb F}[{\mathbb F}{\mathbb F}]_{2n}^A}}
\oint  \frac{d\chi}{2\pi i} \chi^{-6-2n}F_{-4-2n}(\chi) \ \big[G_A^{(1)}(\chi)\big]_{\log|\chi|} = -2n^2-9n-7\,.\la{5.23}
\end{aligned}
\end{equation}
As explained in Section \ref{OPE-sec}, the singlet operators $[{\mathbb F}{\mathbb F}]_{2n}^S$ can mix with $\Phi\Phi$ and two-fermion operators, and 
also $[{\mathbb F}{\mathbb F}]_{2n+1}^A$ can mix with two-fermion states in the same representation. Therefore,
 the corresponding anomalous dimensions 
above should be viewed as averages and more work would be needed to disentangle the mixing.
 The symmetric traceless operators $[{\mathbb F}{\mathbb F}]_{2n}^T$ are not expected to mix, and from 
(\ref{5.23}) we hence get their dimensions to be
\begin{equation}
\la{5.24}
\Delta_{[{\mathbb F}{\mathbb F}]_{2n}^T} = 4+2n-\frac{2 n^2+7 n+5}{\sqrt{\lambda}}+O(\frac{1}{\lambda})\,. 
\end{equation}
Note that $\Delta_{[{\mathbb F}{\mathbb F}]_{2n}^T}$  is 
 the same as $\Delta_{[\Phi {\mathbb F}]_{2n+1}}$  in (\ref{5.14})  and  also 
$\Delta_{[\Phi\Phi]^T_{2n+2}}$ in (\ref{4.36}), indicating that these operators belong to the same supermultiplet.


\section{Circular Wilson loop: comparison to localization}

In the above calculations we assumed the straight Wilson line at the boundary. However, one can map the straight line to the 
circle by a conformal transformation, which allows then to translate correlators of operator insertions on the line to those on the circle. 
 Explicitly, 
we can perform the transformation $t \rightarrow \tan(\tau/2)$, where $-\pi<\tau<\pi$ is the coordinate along the circle. Under 
this transformation, the two-point function of an operator $O_{\Delta}$ inserted in the Wilson loop changes as 
\begin{equation}
\langle \langle O_{\Delta}(t_1)O_{\Delta}(t_2)\rangle \rangle_{\rm line} = \frac{C_{O}}{t_{12}^{2\Delta}} \quad 
\rightarrow 
\quad \langle\langle O_{\Delta}(\tau_1)O_{\Delta}(\tau_2)\rangle \rangle_{\rm circle} = 
\frac{C_{O}}{\big(2\sin\frac{\tau_{12}}{2}\big)^{2\Delta}}\,.
\label{line-to-circle}
\end{equation}
Note that the expectation value of the circular half-BPS Wilson loop is not trivial and given at large $N$ by the well-known expression 
\cite{Erickson:2000af, Drukker:2000rr, Pestun:2007rz}
\begin{equation}
\langle W_{\rm circle}\rangle = \frac{2}{\sqrt{\lambda}} I_1(\sqrt{\lambda})\,,
\label{circle-vev}
\end{equation}
and hence the double-bracket correlator in (\ref{line-to-circle}) requires 
a normalization factor given by this expectation value. 

On the string theory side, 
the transformation from  boundary line to circle simply amounts to changing coordinates on the Euclidean 
 AdS$_2$ worldsheet from the Poincare metric we have been 
assuming above to the hyperbolic disk metric
\begin{equation}
ds^2_2=d\rho^2+\sinh^2\rho\, d\tau^2\,.
\end{equation}
All of our results for the four-point functions of insertions on the line can be then translated to the circle by simply replacing the 
coordinate-dependent prefactors as
\begin{equation}
\frac{1}{t_{12}^{2\Delta_1} t_{34}^{2\Delta_2}}\ \  \rightarrow \ \ 
\frac{1}{(2\sin\frac{\tau_1-\tau_2}{2})^{2\Delta_1} \ (2\sin\frac{\tau_3-\tau_4}{2})^{2\Delta_2}}\,,
\end{equation}
and the conformally invariant cross ratio $\chi$ in \rf{3.2}  is mapped to
\begin{equation}
\chi=\frac{\sin \frac{\tau_1-\tau_2}{2}\ \sin\frac{\tau_3-\tau_4}{2}}{\sin \frac{\tau_1-\tau_3}{2}\ \sin\frac{\tau_2-\tau_4}{2}}\,.
\end{equation}


In the case of the four-point function of $S^5$ fluctuations, it appears to be possible to compare our results to some localization prediction. 
In a series of papers \cite{ Drukker:2007qr, Drukker:2007yx, Giombi:2009ds, Pestun:2009nn, Giombi:2009ek} it was proposed that 
correlation functions in a subsector of supersymmetric Wilson loops and local operators  in ${\cal N}=4$ SYM 
can be computed via localization in terms of 2d YM theory. The relevant Wilson loops, first introduced in \cite{Drukker:2007dw},  
are defined on generic contours on an  $S^2$ subspace of $R^4$ (or $S^4$), and couple to three of the scalar fields in the SYM 
theory, say $\Phi_1,\Phi_2,\Phi_3$, in a way prescribed by supersymmetry:
\begin{equation}
W(C) = {\rm tr} P e^{\oint_C \left(i A_j+\epsilon_{klj}x^k\Phi^{l}\right)dx^j}\,,
\label{eight}
\end{equation}
where $x_i$  
parametrize a unit two-sphere $x_1^2+x_2^2+x_3^2=1$. With such couplings to scalars $\Phi^i$, 
the Wilson loops (\ref{eight}) are 1/8-BPS for generic contour $C$. These operators are mapped under localization\footnote{To be precise, 
this has not yet been proven completely rigorously, as the calculation of the determinant for the fluctuations around the 
localization locus was not computed in \cite{Pestun:2009nn}.} 
to usual Wilson loops in 2d YM on $S^2$. The 1/2-BPS circular Wilson loop is a special case obtained by choosing the contour 
to be a great circle on $S^2$. For instance, taking the equator $x_1=\cos(\tau)$, $x_2=\sin(\tau)$ 
gives the 1/2-BPS operator which couples to $\Phi^3$ only. In this section, we will use this convention for the scalar that couples to 
the 1/2-BPS Wilson loop, to adhere with the definition (\ref{eight}) used in the original papers. 


The relevant local operators appearing in the localization setup are 
chiral primaries with specific position-dependent combination of scalars, which were first studied in \cite{Drukker:2009sf}. 
Recall that a convenient way to write a chiral primary is 
in terms of an auxiliary null 6-vector $\epsilon$ 
\begin{equation}
(\epsilon \cdot \Phi)^J \,,\qquad \epsilon^2=0\,.
\end{equation}
The local operators that are captured by localization are inserted on the $S^2$ and have the form 
\begin{equation}
(x_1\Phi_1+x_2\Phi_2+x_3\Phi_3+i \Phi_4)^J, \qquad \ \ \ \ \  x_1^2+x_2^2+x_3^2=1\ , 
\label{local-twisted}
\end{equation} 
where $x_1,x_2,x_3$ is the point on $S^2$ where the operator is inserted. 
This means that the null 6-vector is position-dependent and given by $\epsilon(x) =(x_1,x_2,x_3,i,0,0)$. These operators are 
mapped by localization \cite{Pestun:2009nn,Giombi:2009ds} 
to powers of the Hodge dual $(i*F)^J$ of the 2d YM field strength, and one can  then compute general mixed correlation functions of 
Wilson loops and local operators using the 2d YM  theory. 

A crucial property of the operators (\ref{local-twisted}) 
is that their correlation functions are position independent, at any coupling \cite{Drukker:2009sf}. From the point of view of 
localization to 2d YM theory, this can be understood as the fact that correlation functions of the field strength dual $*F$ 
are position independent.\footnote{This is because the 2d YM equation of motion is $d*F=0$, and $*F$ is a scalar.} 
In addition to considering correlation functions of Wilson loops with local operators inserted away from the loop, as in \cite{Semenoff:2001xp}, one 
can also insert the local operators (\ref{local-twisted}) along the Wilson loop, which is our main interest here. A calculation of this type was 
carried out in \cite{Bonini:2015fng}, where the calculation in the 2d theory (to the leading order considered there) 
was found to be in agreement with the integrability-based results of \cite{Gromov:2013qga}. 

To make contact with the calculation in Section \ref{4pt-S5}, we should consider the 4-point function of the operators (\ref{local-twisted}) inserted 
along the circular loop
\begin{equation}
\langle \langle \,
\epsilon(\tau_1) \cdot \Phi(\tau_1)\  \epsilon(\tau_2) \cdot \Phi(\tau_2) \ \epsilon(\tau_3) \cdot \Phi(\tau_3) \ \epsilon(\tau_4) \cdot \Phi(\tau_4) \,
\rangle \rangle_{\rm circle}\ , 
\label{4pt-circle}
\end{equation}
where, since the operators are inserted on the great circle in the $(12)$-plane, the null 6-vectors are given by\footnote{
Recall that in this section we are assuming that $\Phi^3$ is the scalar that couples to the Wilson loop, so the $S^5$ fluctuations $y^a$ are 
dual to $\Phi_1,\Phi_2,\Phi_4,\Phi_5,\Phi_6$.}
\begin{equation}
\epsilon(\tau_k) = \big(\cos\,\tau_k,\ \sin\,\tau_k,\ 0,\, i,\, 0,0\big)\equiv \epsilon_k \,,\qquad k=1,\ldots,4\,.
\end{equation}
Let us first check that the two-point function of such operators along the circle is indeed position independent. We have
\begin{equation}
\langle \langle \epsilon(\tau_1)\cdot \Phi(\tau_1)\  \epsilon(\tau_2) \cdot \Phi(\tau_2) \rangle \rangle_{\rm circle}
= C_{\Phi}(\lambda) \frac{\epsilon(\tau_1)\cdot \epsilon(\tau_2) }{\big(2\sin \frac{\tau_{12}}{2}\big)^{2}} = 
-\frac{1}{2} C_{\Phi}(\lambda)\,.
\label{2pt-special}
\end{equation}
As expected, the factor in the numerator coming from the $\tau$-dependent null vector cancels the position dependence of the 
denominator. 

To find (\ref{4pt-circle}), we just have to contract the $SO(5)$ index structures in the result in Section \ref{4pt-S5} with the vectors $\epsilon_k$. 
Note that 
\begin{align} &{\tet
\epsilon_1\cdot \epsilon_2\  \epsilon_3\cdot \epsilon_4 = \big(2\sin \frac{\tau_1-\tau_2}{2}\  2\sin \frac{\tau_3-\tau_4}{2} \big)^2\ }, \qquad
 \frac{\epsilon_1\cdot \epsilon_2 \,\epsilon_3\cdot \epsilon_4}{\epsilon_1\cdot \epsilon_3 \,\epsilon_2\cdot \epsilon_4} = \chi^2 \,,\qquad 
\frac{\epsilon_1\cdot \epsilon_2 \,\epsilon_3\cdot \epsilon_4}{\epsilon_1\cdot \epsilon_4 \,\epsilon_2\cdot \epsilon_3} = \frac{\chi^2}{(1-\chi)^2}\,.
\end{align}
Using these relations  and the decomposition in (\ref{channels}), we find for the unit-normalized connected part of the 4-point function
\begin{equation}
\begin{aligned}
& \frac{\langle \langle 
\epsilon_1 \cdot \Phi(\tau_1) \ \epsilon_2 \cdot \Phi(\tau_2) \ \epsilon_3 \cdot \Phi(\tau_3) \ \epsilon_4 \cdot \Phi(\tau_4)
\rangle \rangle_{\rm circle}^{\rm conn}}{\langle \langle\epsilon\cdot \Phi\epsilon\cdot \Phi\rangle \rangle_{\rm circle}^2} \cr 
=\frac{1}{\sqrt{\lambda}}\Big{[}&G_S^{(1)}(\chi)-\frac{2}{5}G_T^{(1)}(\chi)
+\frac{1}{\chi^2}\big(G_T^{(1)}(\chi)+G_A^{(1)}(\chi)\big) 
+\frac{(1-\chi)^2}{\chi^2}\big(G_T^{(1)}(\chi)-G_A^{(1)}(\chi)\big)\Big{]}\ . 
\end{aligned}
\end{equation}
Plugging in the explicit functions of cross-ratio from (\ref{GSTA}), one can verify that the position dependence completely cancels out 
and we end up with 
\begin{equation}
\frac{\langle \langle 
\epsilon_1 \cdot \Phi(\tau_1)\ \epsilon_2 \cdot \Phi(\tau_2)\ \epsilon_3 \cdot \Phi(\tau_3) \ \epsilon_4 \cdot \Phi(\tau_4)
\rangle \rangle_{\rm circle}^{\rm conn}}{\langle \langle\epsilon\cdot \Phi\ \epsilon\cdot \Phi\rangle \rangle_{\rm circle}^2}  = -\frac{3}{\sqrt{\lambda}}+O(\frac{1}{\lambda})\,.
\label{4pt-special}
\end{equation}

Let us now compare this result with the prediction of localization. One should compute the 4-point function 
$\langle \langle \tilde F(\tau_1)\tilde F(\tau_2)\tilde F(\tau_3)\tilde F(\tau_4)\rangle \rangle_{\rm YM_2}$ in 2d YM, where we introduced for convenience 
the shorthand $\tilde F \equiv i*F$ for the dual of the field strength, which is inserted four times along the circular Wilson loop. 
A shortcut to this calculation may be obtained by starting from a more general contour $C$ and noticing 
that insertions of $i*F$ are equivalent to taking derivatives of the Wilson loop expectation 
value with respect to the area.\footnote{In general, under a small deformation of the contour, the Wilson loop expectation value 
varies as $\langle W(C+\delta C)\rangle = \langle {\rm tr}P(1+\int d\tau \delta x^{\mu} \dot x^{\nu}iF_{\mu\nu}+\ldots)e^{ \int  i A}\rangle$. 
In 2d, we can write  
$\int d\tau \delta x^{\mu} \dot x^{\nu}iF_{\mu\nu}=\int d\tau \delta x^{\mu} \dot x^{\nu}\sqrt{g}\epsilon_{\mu\nu} i*F$. The 
factor $\int d\tau \delta x^{\mu} \dot x^{\nu}\sqrt{g}\epsilon_{\mu\nu}$ measures the change in area of the Wilson loop.} For a general contour $C$ singling 
out areas $A_1$, $A_2$ on $S^2$, with $A_1+A_2=4\pi$ (we take unit radius), the invariance under area preserving diffeomorphisms of 2d YM implies that 
the expectation value is given by the same expression (\ref{circle-vev}) up to an area-dependent rescaling of the coupling
\begin{equation}
\langle W_{A_1}\rangle  = \frac{2}{\sqrt{\lambda'}} I_1(\sqrt{\lambda'})\,,\qquad\qquad  \lambda' \equiv \frac{A_1A_2}{4\pi^2}\lambda=\frac{A_1(4\pi-A_1)}{4\pi^2}\lambda\,.
\end{equation}
This is the expectation value of the general 1/8-BPS operator (\ref{eight}). The 1/2-BPS circle corresponds to the special case $A_1=2\pi$    when $\l'=\l$. Then, taking 
derivatives of $\log\langle W_{A_1}\rangle$ with respect to $A_1$ and setting $A_1=2\pi$ 
after that yields the connected correlators of $i*F$ inserted along 
the circle. 

For instance, the two-point function is given by
\begin{equation}
\langle \langle \tilde F(\tau_1)\tilde F(\tau_2)\rangle \rangle_{\rm YM_2} = 
\frac{\partial^2}{\partial A_1^2}\log\langle W_{A_1}\rangle\Big|_{A_1=2\pi} = 
-\frac{\sqrt{\lambda } I_2(\sqrt{\lambda })}{4 \pi ^2 I_1(\sqrt{\lambda })}\,.
\end{equation}
This implies that in (\ref{2pt-special}) \ $C_{\Phi}(\lambda)=\frac{\sqrt{\lambda } I_2(\sqrt{\lambda })}{2\pi ^2 I_1(\sqrt{\lambda })}$, \ 
in agreement with the Brehmsstrahlung function of \cite{Correa:2012at}. 

For the connected 4-point function, we get
\begin{equation}
\frac{\langle \langle \tilde F(\tau_1)\tilde F(\tau_2)\tilde F(\tau_3)\tilde F(\tau_4)\rangle \rangle_{\rm YM_2}^{\rm conn}}
{\langle \langle\tilde F \tilde F\rangle \rangle_{\rm YM_2}^2}
=\frac{\frac{\partial^4}{\partial A_1^4}\log\langle W_{A_1}\rangle\big|_{A_1=2\pi}}{\left(\frac{\partial^2}{\partial A_1^2}\log\langle W_{A_1}\rangle\big|_{A_1=2\pi}\right)^2}
=\frac{3 (\lambda +4) \big[I_1(\sqrt{\lambda })\big]^2-3 \lambda \big[ I_0(\sqrt{\lambda })\big]^2}{\lambda  \big[I_2(\sqrt{\lambda })\big]^2}\,.
\end{equation} 
Expanding at large $\lambda$, this gives 
\begin{equation}
\frac{3 (\lambda +4) \big[I_1(\sqrt{\lambda })\big]^2-3 \lambda \big[ I_0(\sqrt{\lambda })\big]^2}{\lambda  \big[I_2(\sqrt{\lambda })\big]^2} = 
-\frac{3}{\sqrt{\lambda}}+\frac{45}{8 \lambda^{3/2}}+\ldots \ , 
\end{equation}
and we see that the leading term agrees with our result (\ref{4pt-special}) coming from tree-level connected diagrams in AdS$_2$. 

\section*{Acknowledgments}
We are grateful to Nadav Drukker, Vladimir Kirilin, Juan Maldacena, Marco Meineri, Carlo Meneghelli, Eric Perlmutter and Vasily Pestun for useful discussions. 
RR acknowledges the hospitality of KITP at UC Santa Barbara in the program   ``Scattering amplitudes and beyond", 
during the final stages of this work.
The work of SG is supported in part by the US NSF under Grant No.~PHY-1620542.
The work of RR is supported in part by the US DOE under Grant DE-SC0013699 and while at KITP also  by the 
US NSF under Grant No.~PHY11-25915.
The work of AAT  was   supported by the ERC Advanced grant no. 290456,
 the  STFC Consolidated grant ST/L00044X/1
  and   the Russian Science Foundation grant 14-42-00047 at Lebedev Institute.

\begin{appendices}
\section{Toy model: scalar in AdS$_2$ with $\varphi^4$ interaction}

As a simple toy model, let us consider a scalar in AdS$_2$ with a simple quartic self-interaction
\begin{equation}\
S = \int d^2x\sqrt{g}\Big(\frac{1}{2}g^{\mu\nu}\partial_{\mu}\varphi \partial_{\nu}\varphi 
+ \frac{1}{2}m^2\varphi^2+\frac{g}{4!}\varphi^4\Big)\,, 
\end{equation}
where we assume the Poincare metric $ds^2 = {1\ov z^2} (dz^2+dt^2)$. 
Tree level Witten diagrams obtained from this model yield conformally invariant correlation functions of an operator $O(t)$ 
at the boundary with scaling dimension given by
$\Delta(\Delta-1)=m^2$. The tree-level 4-point function is straightforward to compute
\begin{equation}
\langle O(t_1)O(t_2)O(t_3)O(t_4)\rangle = -g {\cal C}_{\Delta}^4 D_{\Delta\Delta\Delta\Delta}(t_1,t_2,t_3,t_4) 
=-g\frac{{\cal C}_{\Delta}^4\sqrt{\pi}\Gamma(2\Delta-\ha)}{2[\Gamma(\Delta)]^4}\frac{1}{t_{12}^{2\Delta}t_{34}^{2\Delta}} 
\chi^{2\Delta}\bar{D}_{\Delta\Delta\Delta\Delta}(\chi)
\end{equation}
Specializing to the case of a massless scalar, so that $\Delta=1$, this  may be written as (cf. \rf{4.8})
\begin{equation}
\langle O(t_1)O(t_2)O(t_3)O(t_4)\rangle =-\frac{g}{4\pi}\frac{({\cal C}_{\Delta=1})^2}{t_{12}^2t_{34}^2}\chi^2 \bar{D}_{1111}(\chi)\ , 
\end{equation}
with $\bar{D}_{1111}(\chi)$ given in (\ref{Db1-1d}). From this result we can extract the anomalous dimension of the $[OO]_{2n}\sim O\partial_t^{2n}O$ 
operators as explained in the main text. The leading order OPE coefficients are given by (\ref{OPE-gen-free1}) with $\Delta=1$ 
\begin{equation}
c_{OO[OO]_{2n}}^{(0)} = \frac{2\big[ \Gamma (2 n+2)\big]^2\,  \Gamma (2 n+3)}{\Gamma (2 n+1) \, \Gamma (4 n+3)}\ , 
\end{equation}
and extracting the 
coefficient of $\log(\chi)$ in $\bar{D}_{1111}(\chi)$, the anomalous dimensions are given by 
\begin{equation}
\gamma_{[OO]_{2n}}^{(1)} = \frac{1}{c_{OO[OO]_{2n}}^{(0)}} 
\oint  \frac{d\chi}{2\pi i} \chi^{-3-2n}F_{-1-2n}(\chi) \frac{g \chi ^2}{2 \pi  (1-\chi)}\ , 
\end{equation}
which yields
\begin{equation}\la{a6}
\Delta_{[OO]_{2n}} = 2+2n+\frac{g}{4 \pi} \frac{1}{(2 n+1) (n+1)}+O(g^2)\ . 
\end{equation}
Note that unlike the results we obtained above  from the worldsheet model \rf{2.4}, the anomalous dimensions in \rf{a6}are positive and also they go to zero at large $n$. 

One can similarly consider the case of a $m^2=2$ scalar, i.e. $\Delta=2$. Then we get 
\begin{equation}
\begin{aligned}
&\langle O(t_1)O(t_2)O(t_3)O(t_4)\rangle =-\frac{5g}{12\pi}\frac{({\cal C}_{\Delta=2})^2}{t_{12}^4t_{34}^4}\chi^4 \bar{D}_{2222}(\chi)\ , \\
&\bar{D}_{2222}(\chi) = \frac{\chi  (2 \chi -5)+5}{30 (\chi -1)^3}\log(\chi^2)
-\frac{2 \chi ^2+\chi +2}{30 \chi ^3}\log\big((1-\chi)^2\big)-\frac{2 ((\chi -1) \chi +1)}{15 (\chi -1)^2 \chi ^2}\ . 
\end{aligned}
\end{equation}
The leading OPE coefficients are obtained from (\ref{OPE-gen-free1}) with $\Delta=2$
\begin{equation}
c_{OO[OO]_{2n}}^{(0)} =\frac{\big[\Gamma (2 n+4)\big]^2\  \Gamma (2 n+7)}{18\,  \Gamma (2 n+1)\  \Gamma (4 n+7)}\ , 
\end{equation}
and the anomalous dimensions are then given by
\begin{equation}
\gamma_{[OO]_{2n}}^{(1)} = \frac{1}{c_{OO[OO]_{2n}}^{(0)}} 
\oint  \frac{d\chi}{2\pi i} \chi^{-5-2n}F_{-3-2n}(\chi)\  \frac{g \chi ^4 ((5-2 \chi ) \chi -5)}{36 \pi  (\chi -1)^3}\ . 
\end{equation}
This yields the result
\begin{equation}
\Delta_{[OO]_{2n}}=4+2n +\frac{g}{4\pi} \frac{(n+1) (2 n+5)}{(n+2) (n+3) (2 n+1) (2 n+3)}+O(g^2)\,.
\end{equation}

\end{appendices}

\bibliographystyle{ssg}
\bibliography{Wilsonbib}
\end{document}